\documentclass[10pt,a4paper]{article}
\usepackage[T1]{fontenc}
\usepackage{inputenc}
\usepackage[english]{babel}
\usepackage[nohead,includefoot,margin=2cm]{geometry}
\usepackage[compact,raggedright,small]{titlesec}
\usepackage[affil-it]{authblk}
\usepackage[runin]{abstract}
\usepackage{multicol}
\usepackage{graphicx}
\usepackage{mathptmx}
\usepackage{latexsym}
\usepackage{xcolor}
\newcommand*{\pacsname}{PACS numbers}
\newenvironment{pacs}{%
  \begin{list}{}{%
    \settowidth{\labelwidth}{\pacsname:}
    \setlength{\leftmargin}{\labelwidth}
    \addtolength{\leftmargin}{\labelsep}}
  \item[\pacsname:]}
{\end{list}}
\addto{\Affilfont}{\small}

\title{\large\bfseries A New  Fate of a Warped 5D FLRW Model with a U(1) Scalar Gauge Field}
\date{\small (Accepted:  Dec., 2015;  Found. of Phys.)}
\author[1]{R. J. Slagter}
\author[2]{S. Pan}
\affil[1]{ASFYON, Astronomisch Fysisch Onderzoek Nederland   and University of Amsterdam, department of physics.  1405EP Bussum, The Netherlands. email: info@asfyon.com}
\affil[2]{University of Jadavpur, department of mathematics,  Kolkata-700032, India.  email: span@research.jdvu.ac.in}
\begin{document}
\maketitle
\begin{abstract}
If we live on the weak brane with zero effective cosmological constant in a warped 5D bulk spacetime, gravitational waves and brane fluctuations can be generated by a part of the 5D  Weyl tensor and carries information of the gravitational field outside the brane. We consider on a cylindrical symmetric warped FLRW background a U(1) self-gravitating  scalar field coupled to a gauge field without bulk matter. It turns out that brane fluctuations  can be formed dynamically, due to the modified energy-momentum tensor components of the scalar-gauge field ("cosmic string"). As a result, we find that the late-time behavior could be  significantly deviate from the standard evolution of the universe. The effect is triggered by the time-dependent warpfactor with two branches of the form \\
$\frac{\pm1}{\sqrt{\tau r}}\sqrt{(c_1e^{\sqrt{2\tau} t}+c_2e^{-\sqrt{2\tau} t})(c_3e^{\sqrt{2\tau} r}+c_4e^{-\sqrt{2\tau} r})}$ ( with $\tau , c_i$ constants) and the modified brane equations comparable with a dark energy effect. This is a brane-world mechanism, not present in standard 4D FLRW, where the large disturbances are rapidly damped as the expansion proceed. Because gravity can propagate in the bulk, the cosmic string can build up a huge angle deficit (or mass per unit length) by the warpfactor and can induce massive KK-modes felt on the brane. Disturbances in the spatial components of the stress-energy tensor cause cylindrical symmetric waves, amplified due to the presence of the bulk space and warpfactor. They could survive the natural damping due to the expansion of the universe. It turns out that one of the metric components  becomes singular at the moment the warp factor develops a extremum. This behavior could have influence on the possibility of a transition from acceleration to  deceleration or vice versa.
\end{abstract}

\section{Introduction}
It is conjectured that the expansion of our universe is accelerating. Recent observations provide strong evidence of this acceleration. The explanation of this remarkable phenomenon is rather difficult: one needs a dark energy field with an effectively negative pressure, $p<-\frac{1}{3}\rho$. From several independent observational data, one finds $\Omega_\Lambda \approx 0.7$ and $\Omega_M\approx 0.3$, where $\Omega_M$ and $\Omega_\Lambda$ stand for the energy densities of matter and dark energy respectively with respect to the critical ones \cite{Sperg:2003,Peir:2003,Sperg1:2007,Per:1999,Riess:2006}. We should live now in the cosmological constant dominated era (and approximately flat, $\Omega_0= \Omega_{M0}+ \Omega_{\Lambda0} \approx 1$), while at earlier times there was a radiation and matter dominated era.

There are some fundamental questions. First, will the acceleration last forever. Secondly, why there is the huge discrepancy between $\rho_\Lambda \simeq 10^{-123}$ in Planck units and vacuum energy density $\langle\rho_V \rangle \simeq 10^{-3}$ which is $10^{120}$ times greater than the value we need to accelerate the expansion of the universe. Thirdly, the value must be incredibly fine-tuned, $\Omega_\Lambda \sim \Omega_M$. So it would be a logical step to try to explain the late-time acceleration without the need for dark energy \cite{Roy:2006}. It is also conjectured that one needs an inflaton field in the very early stage of our universe, to solve the flatness and horizon problem in the standard model of cosmology. This is the inflationary cold dark matter model with cosmological constant ($\Lambda$CDM). It could also predict the existence of fluctuations we observe in the CMB shortly before the end of inflation. The inflaton-field could be the well-known scalar-Higgs field with the mexican-hat potential. This model has lived up to its reputation. It was successful in the explanation of superconductivity, i.e., the Ginsburg-Landau theory, in the standard model of particle physics, in the general relativistic solution of the self-gravitating Nielsen-Olesen vortex (cosmic string) \cite{Felsager:1981,Nielsen:1973,Garf:1985}  and could play a fundamental role in warped spacetimes \cite{Slag1:2012,Slag2:2014,Slag3:2014}.
Cosmic strings are U(1) scalar-gauge vortex solutions in general relativity (GR)\cite{Garf:1985,Lag1:1987,Lag2:1989}. It is conjectured that in any field theory which admits cosmic strings, a network of strings inevitably forms at some point during the early universe. However it is doubtful if they persist to the present time. Evidence of these objects would give us information at very high energies in the early stages of the universe. From recent observations by COBE, WMAP and Planck satellites, it was concluded that cosmic strings cannot provide satisfactory explanation for the magnitude of the initial density perturbations from which galaxies and clusters grew. The interest in cosmic strings faded away, mainly because of the inconsistencies with the power spectrum of CMB.

Cosmological cosmic strings can also be investigated on a  Friedmann-Lemaitre-Robertson-Walker (FLRW) background. However, the string-cosmology spacetime essentially looks like a scaled version of a string in a vacuum spacetime and the corrections in the field equations due to cylindrical gravitational radiation are rapidly damped and are negligible in any physical regime \cite{Greg:1989}. The reason of this result comes from the notion that the string radius $r_{CS}$ is much smaller than the Hubble radius $R_H$ in the post-inflationary era. The ratio $r_{CS}/R_H = \partial_t C /C $, where $C$ is the factored out overall scale factor in the metric, is then of the order $10^{-57}$. The resulting field equations look like the generalized Nielsen-Olesen vortex system. Only in the pre-inflationary epoch, radiative corrections could be very large. While the inflaton field plays a crucial role in the early stage of our universe, it could play a comparable role at much later times, if we modify gravity by considering warped brane-world models. It could be possible that there exists a correlation between the accelerating universe and large extra dimensions in brane-world models.

When it was realized that cosmic strings could be produced within the framework of superstring theory inspired cosmological models, a revival of cosmic strings occurred. These so-called cosmic superstrings can play the role of cosmic strings in the framework of string theory or M-theory, i.e., brane-world models. Supersymmetric GUT's can even demand the existence of cosmic strings. Physicists speculate that extra spatial dimensions could exist in addition to our ordinary 4-dimensional spacetime. The idea that spacetime could have more than four dimensions was first proposed by Kaluza and Klein (KK) in the early 20$^{th}$ century \cite{Kal:1921,Klein:1926}.
These theories can be used to explain several of the shortcomings of the Standard Model, i.e., the unknown origin of dark energy and dark matter and the weakness of gravity (hierarchy problem). In these models, the weakness of gravity might be fundamental. One might naively imagine that these extra dimensions must be very small, i.e., curled up and never observable.
Super-massive strings with $G\mu >1$, could be produced when the universe underwent phase transitions at energies much higher than the GUT scale. Recently there is growing interest in the so-called brane-world models, first proposed by Arkani-Hamed, Dimopoulos and Dvali (ADD) \cite{ADD:1998,ADD1:1999} and I. Antoniadis et al. \cite{IA:1998}, which was extended by Randall and Sundrum (RS) \cite{RS:1999}. In these models, the extra dimension can be very large compared to the ones predicted in string theory, i.e., of order of millimeters. The difference with the standard superstring model is that the compactification rely on the curvature of the bulk. The huge discrepancy between the electro-weak scale, $M_{EW}=10^3 GeV$ and the gravitational mass scale $M_{Pl}=10^{19} GeV$ will be suppressed by the volume of the extra dimension $y$, or the curvature in that region. This effect can also be achieved in the RS models by a warpfactor.
In the RS-2 model, there are two branes, the visible and the gravity brane at $y= 0$. The branes have equal and opposite tensions $\pm \Lambda_4$.
At low energy, a negative bulk cosmological constant will prevent gravity to leak into the extra dimension, $\Lambda_5 = -6/R_0^2=-6/\mu^2$, with $\mu$ the corresponding energy scale and $R_0$ the compactification radius. The $\Lambda_5$ squeeze the gravitational field closer to the brane at $y= 0$. In the RS--1 model, one pushes the negative tension brane $y\rightarrow \infty$. If one fine-tunes the $\Lambda_4 =3M_{Pl}^2/ 4\pi R_0^2$, then this ensures a zero effective cosmological constant on the brane.  The infinite extra dimension makes, however, a finite contribution to the 5D volume due to the warpfactor. Because of the finite separation of the branes in the RS-2 model, one obtains so-called effective 4D modes (KK-modes) of the perturbative 5D graviton on the 4D brane. These KK-modes will be massive from the brane viewpoint. In the RS-1 model, the discrete spectrum disappears and will form a continuous spectrum \cite{maart:2003,maart:2010,shir:2000}.

It will be clear that compact objects, such as black holes and cosmic strings, could have tremendous mass in the bulk, while their warped manifestations in the brane can be consistent with observations. So, brane-world models could overcome the observational bounds one encounters in cosmic string models. $G\mu$ could be warped down to GUT scale, even if its value was at the Planck scale. Although static solutions of the U(1) gauge string on a warped spacetime show significant deviation from the classical solution in 4D \cite{slag3:2012}, one is interested in the dynamical evolution of the effective brane equations. Wavelike disturbances triggered by the huge mass of the cosmic string in the bulk, could have observational effects in the brane. One conjectures that these disturbances could act as an effective dark energy field. The question is if the pulse-like cylindrical waves have the desired asymptotic behavior at null infinity.


Here, we will investigate the late-time evolution of a warped 5D FLRW model when a U(1) scalar field coupled to a gauge field is present. In section II, we outline the model and in section III, we present some numerical solutions of the model.
Finally, we have summarized our results in section IV.
\section{The Field Equations}

In order to embed a cosmological cosmic string in a FLRW model, one has to modify the metric. The spacetime of a cosmic string is invariant under boosts in the $z$ direction,
whereas the FLRW spacetime is not. Gregory\cite{Greg:1989} pointed out that in the static case, if the radius of the string core is small compared to the Hubble radius $R_H$, the embedding is easily done.
For a non-static radiating cylindrically symmetric spacetime, one can analyze the behavior of gravitational waves generated by the local strings on the expansion of the universe if we consider the zero-thickness limit of non-singular spacetimes containing a cylindrical distribution of stress-energy embedded on a cosmological background.

The importance of cylindrical symmetric gravitational waves was first noticed by Beck \cite{Beck:1925} and rediscovered by  Einstein and Rosen (ER)\cite{Ros:1937}. A substantial part of the present knowledge on gravitational waves originated from their wavelike solutions of the fully non-linear gravitational field equations.
However, these original ER cylindrical wave solutions of the Einstein's vacuum field equations inhabit a universe that is conical rather than flat at infinite cylindrical radius\cite{Weber:1957}. To overcome this problem, one could impose artificially non-conicality constraints\cite{Gow:2007,Gow2:2007} in order to obtain a Kasner-like spacetime. In a warped 5-dimensional brane-world model with a U(1) gauge cosmic string, these constraints are superfluous because the effective brane spacetime is non-conical\cite{Slag1:2012}.
Much insight into wavelike solutions on a cylindrically spacetime can also be obtained from well known stationary axially symmetric solutions, like the Weyl, Papapetrou, Lewis-van Stockum, and Kerr solution, through a complex transformation $z\rightarrow it, t\rightarrow iz$ \cite{Step:2009}. The cylindrical symmetric metric is (with two killing vectors $\partial / \partial z, \partial / \partial\varphi$, both spacelike and hypersurface orthogonal)
\begin{equation}
ds^2=e^{2(\gamma(t,r)-\psi(t,r))}(-dt^2+dr^2)+e^{2\psi(t,r)}dz^2+r^2e^{-2\psi(t,r)}d\varphi^2. \label{eqn1}
\end{equation}
Without any matter source, one then obtains a wave equation for $\psi$, which is decoupled from $\gamma$. Once $\psi$ is solved, $\gamma$ can be solved by quadratures and the behavior of $\psi$ and $\gamma$ at null infinity is well known, i.e., the Einstein-Rosen pulse waves. One can construct incoming and outgoing radiation solutions \cite{Stach:1968,Ash:1996} in terms of the retarded and advanced coordinates $u= t-r$ and $v= t+r$. For a non-vacuum situation, the equations will not decouple.

It was found by Gregory \cite{Greg:1989} that a U(1) cosmic string can be embedded into a flat FLRW spacetime along the polar axis, when the Hubble radius is much larger than the string-core. The approximated spacetime is
\begin{equation}
ds^2=a(t)^2\Bigl[-dt^2+dr^2+K(r)^2dz^2+(1-4\pi G\mu)^2S(r)^2d\varphi^2\Bigr], \label{eqn2}
\end{equation}
with $\mu$ the mass per unit length of the string, or angle deficit and $K(r)=\cos r, 1, \cosh r$ and $ S(r)=\sin r, r, \sinh r$ when $k=1, 0, -1 $ respectively. One can match this spacetime on the  well-known FLRW spacetime by a  transformation from $(R,\theta)$ to $(r,z)$ by defining  $S(r)=R\sin \theta$ and $T(z)=(1-kR^2)^{-\frac{1}{2}} $, with $T(z)=\tan z, z,\tanh z$ when $k=1,0$ or -1 respectively \cite{And:2003}. One should also like to impose matching conditions across the boundary $r_{CS}$ of the interior and exterior spacetime (although for radiating strings, this will be difficult). For late-time solutions, it is not necessary to consider the matching conditions between the interior string solutions and the exterior FLRW spacetime.
It turns out that when the width of the strings $r_{CS}$ is smaller than the Hubble radius, the disturbances are neglectable \cite{Greg:1989}.

On a warped spacetime, where the warpfactor has a significant effect on the cosmic string solution \cite{Slag1:2012,Slag2:2014}, disturbances  can be significant. In the bulk, the mass of the cosmic string will be huge, inducing a back reaction on the brane equations.

\subsection{The Bulk Equations}
Following Shiromizu et al. \cite{Shir:2000}, we have 5-dimensional Einstein equations with a bulk cosmological constant $\Lambda_5$ and brane energy-momentum as source
\begin{equation}
{^{(5)}G}_{\mu\nu}=-\Lambda_5{^{(5)}g}_{\mu\nu}+\kappa_5^2 \delta(y)\Bigl(-\Lambda_4 {^{(4)}g}_{\mu\nu}+{^{(4)}T}_{\mu\nu}\Bigr), \label{eqn3}
\end{equation}
with $\kappa_5= 8\pi {^{(5)}G}= 8\pi/{^{(5)}M}_{pl}^3$, $\Lambda_4$ the brane tension, $x^\mu =(t,x^i,y)$, ${^{(4)}g}_{\mu\nu}={^{(5)}g}_{\mu\nu}-n_\mu n_\nu$,
and $n^\mu$ the unit normal to the brane. The ${^{(5)}M}_{pl}$ is the fundamental 5D Planck mass, which is much smaller than the effective Planck mass on the brane, $\sim 10^{19}$ GeV. We consider here the matter field ${^{(4)}T}_{\mu\nu}$ confined to the brane, i.e., the U(1) scalar-gauge field, written,  in polar coordinates $(t,r,z,\varphi )$, in the form \cite{Garf:1985}
\begin{eqnarray}
\Phi=\eta X(t, r)e^{i\varphi},\quad A_\mu =\frac{1}{ \epsilon }\bigl[P(t, r)-1\Bigr]\nabla_\mu\varphi, \label{eqn4}
\end{eqnarray}
with $\eta$, the vacuum expectation value of the scalar field, $\epsilon$ the coupling constant and $r=\sqrt{(x^i)^2}$  the radial coordinate. Further, we take for the potential of the scalar field the well-known "Mexican-hat" form
$V(\Phi)=\frac{1}{8}\beta(\Phi^2-\eta^2)^2 $.
Let us consider the cylindrically symmetric warped spacetime
\begin{equation}
ds^2 = {\cal W}(t, r, y)^2\Bigl[e^{2(\gamma(t, r)-\psi(t, r))}(-dt^2+ dr^2)+e^{2\psi(t, r)}dz^2+ r^2 e^{-2\psi(t, r)}d\varphi^2\Bigr]+ dy^2, \label{eqn5}
\end{equation}
with ${\cal W}$ be the warpfactor and $y$ be the bulk space coordinate.\\
\begin{figure}[pb]
\centerline{
\includegraphics[width=4.5cm]{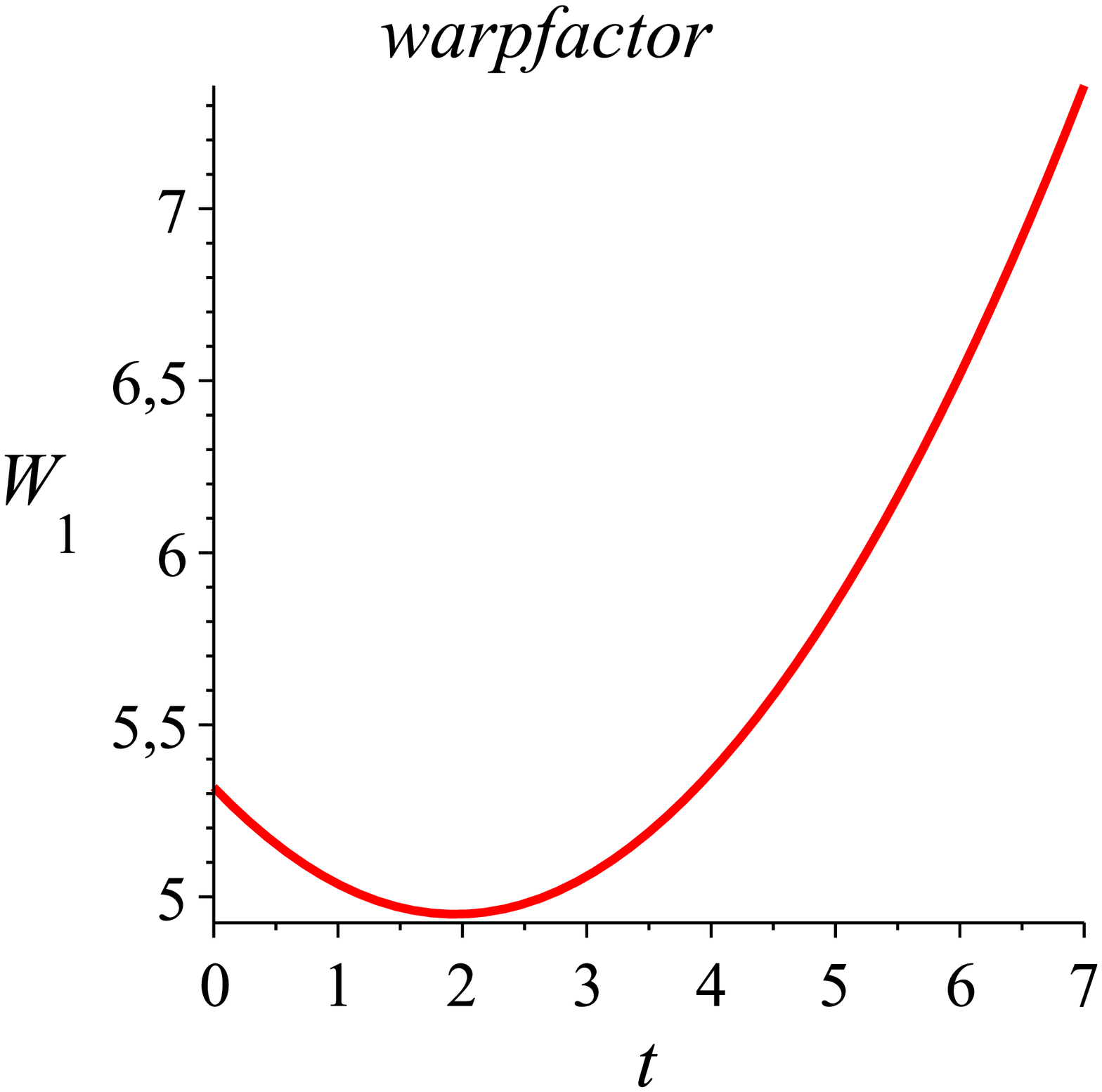}
\includegraphics[width=4.5cm]{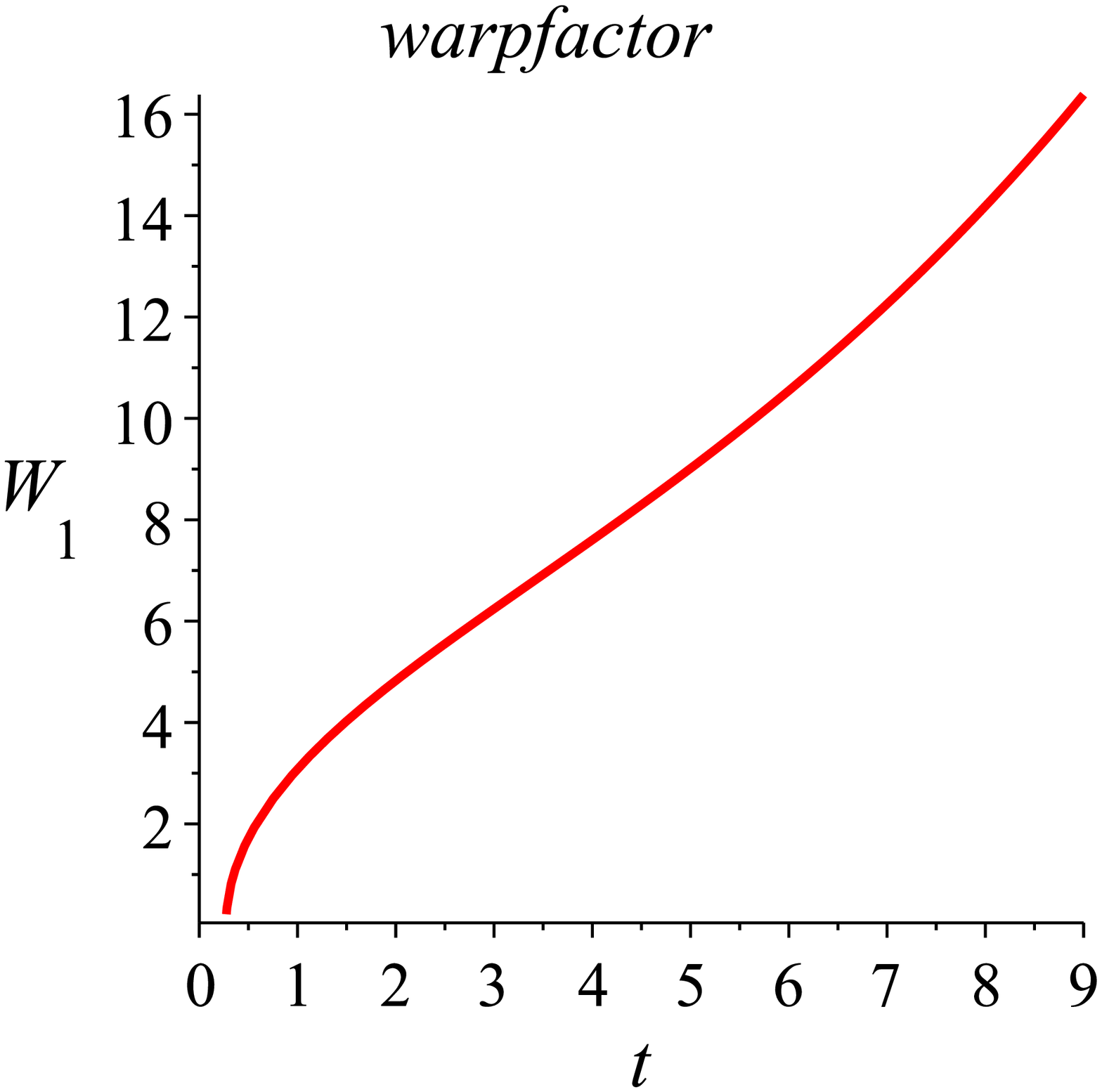}
\includegraphics[width=6.cm]{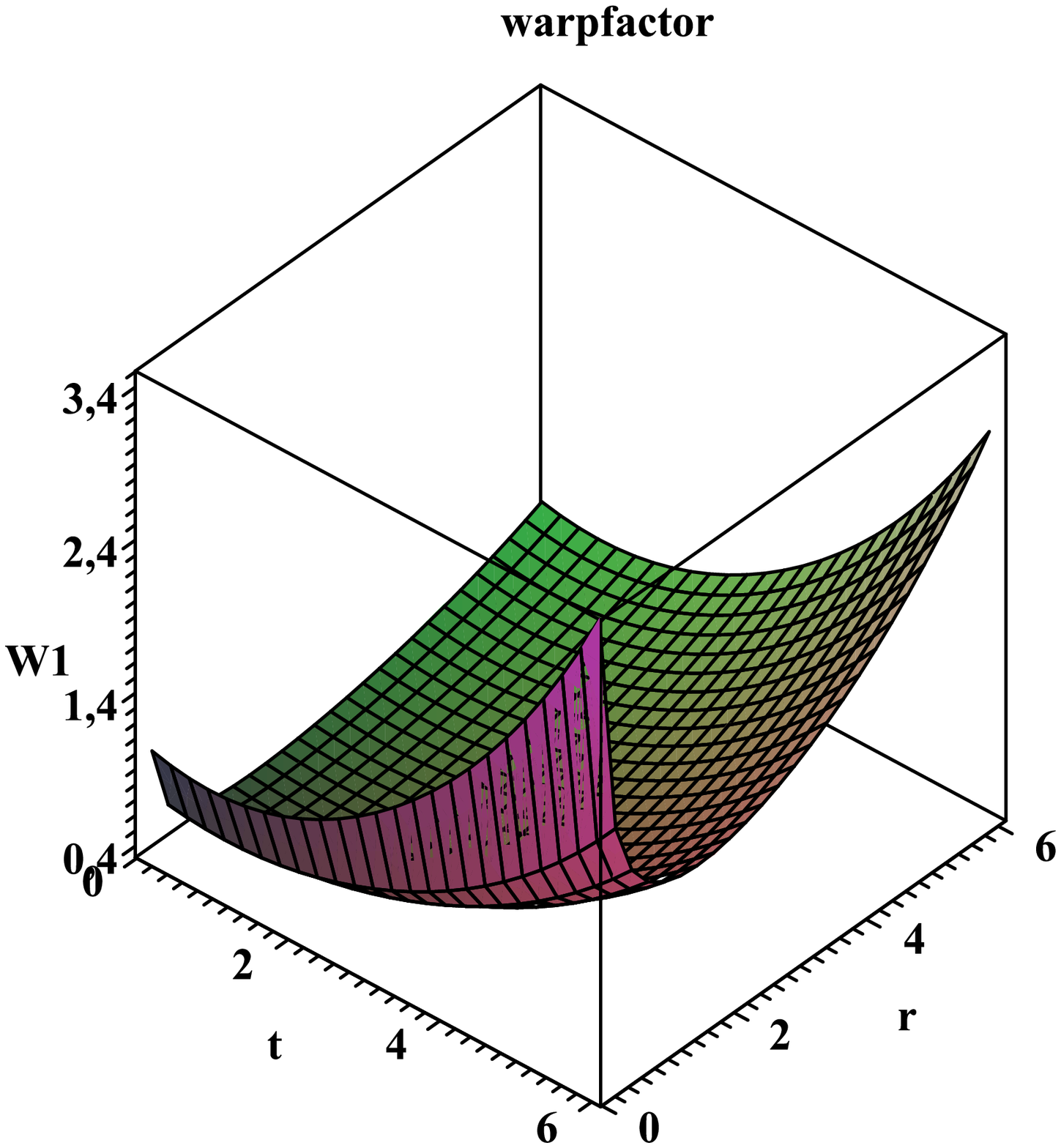}}
\vskip 1.2cm
\caption{Typical solutions of the warpfactor $W_1$. Left the two different possibilities for the time-dependent part  for suitable values of the constant $d_1$ and $d_2$ : a minimum or an inflection point. Right a 3D plot for some values of $\tau $ and $d_i$ [see Eq. (\ref{eqn10})].}
\end{figure}

From the 5D equations we obtain from the $(t, y)$ component of the 5D Einstein equations (a dot means $\frac{\partial}{\partial t}$)
\begin{equation}
\partial_y\dot{{\cal W}}=\frac{\dot{{\cal W}}\partial_y{\cal W}}{{\cal W}}. \label{eqn6}
\end{equation}
If we write  ${\cal W}(t,r,y)=W_1(t,r)W_2(y)$ as particular solution of Eq. (\ref{eqn6})  and substitute this expression into the 5D Einstein equations,
we can then separate the y-dependent parts containing $W_2$:
\begin{eqnarray}
\partial_{yy}W_2=-\frac{(\partial_y W_2)^2}{W_2}-\frac{1}{3}\Lambda_5W_2-\frac{c_1}{W_2}, \quad (\partial_y W_2)^2=-\frac{1}{6}\Lambda_5W_2^2+c_2, \label{eqn7}
\end{eqnarray}
with a simplified solution (for the moment, $c_1, c_2 =0$)
\begin{equation}
W_2(y)=e^{\sqrt{- \frac{1}{6} \Lambda_5}(y- y_0)}. \label{eqn8}
\end{equation}
We see that only a negative bulk cosmological constant make sense. For $c_1, c_2 \neq 0$ one obtains a solution with positive bulk cosmological constant.
The y-independent part of the 5D Einstein equation for $W_1(t,r)$ becomes
\begin{equation}
\ddot{W_1}=W_1''+\frac{1}{W_1}(W_1'^2-\dot{W_1}^2)+\frac{2}{r}W_1', \label{eqn9}
\end{equation}
with a prime $\frac{\partial}{\partial r}$. A typical solution is
\begin{equation}
W_1(t,r)=\frac{\pm1}{\sqrt{\tau r}} \sqrt{\Bigl(d_1 e^{(\sqrt{2\tau})t}-d_2e^{-(\sqrt{2\tau})t}\Bigr)\Bigl(d_3 e^{(\sqrt{2\tau})r}-d_4e^{-(\sqrt{2\tau})r}\Bigr)}, \label{eqn10}
\end{equation}
with $\tau, d_i$ some constants. So we have two branches, i.e., the plus and minus sign in Eq. (\ref{eqn10}). Figure 1 represents the typical plots of $W_1(t, r)$ for different sets of the constants $\tau$ and $d_i$. In general, $W_1$ can possess a saddle-point or  extremal values, as can be seen from the time-dependent part of $W_1$ of Figure 1. We shall see in section (2.2) that the appearance of a minimum ( or maximum in case of  the other branch) will make the equation for $\dot\gamma$ divergent. This could have a significant influence on a transition from acceleration to deceleration or vice versa.
We shall also see in Section 2.2 that these solutions for the warpfactor are consistent with the supplementary equations following from the 5D Einstein  and the Bianchi equations. It turns out that the equations for $W_1(t,r)$ can not be isolated from the effective 4D brane equations, indicating that $W_1$ is a really a warpfactor-effect.

\subsection{The Effective Brane Equations}
The field equations induced on the brane can be derived using the Gauss--Codazzi equations together with the Israel--Darmois junction conditions at the brane and the $Z_2$ symmetry \cite{Sas:2000}. The modified Einstein equations become
\begin{equation}
{^{(4)}G}_{\mu\nu}=-\Lambda_{eff}{^{(4)}g}_{\mu\nu}+\kappa_4^2 {^{(4)}T}_{\mu\nu}+\kappa_5^4{\cal S}_{\mu\nu}-{\cal E}_{\mu\nu}, \label{eqn11}
\end{equation}
where $\Lambda_{eff}=\frac{1}{2}(\Lambda_5+\kappa_4^2\Lambda_4)=\frac{1}{2}(\Lambda_5+\frac{1}{6}\kappa_5^4\Lambda_4^2)$ and $\Lambda_4$ is the vacuum energy in the brane (brane tension).  The latter equality sign is a consequence of the relation between the 4- and 5-dimensional Planck mass in the braneworld approach, $\kappa_5^4=6\frac{\kappa_4^2}{\Lambda_4}$\cite{Dur:2005,Shir:2000}. If in addition the brane tension is related to the 5-dimensional coupling constant and the cosmological constant by $ \frac{1}{6}\Lambda_4^2\kappa_5^4=-\Lambda_5$, then $\Lambda_{eff}=0$ and we are dealing with the RS-fine tuning condition\cite{RS:1999}

The first correction term ${\cal S}_{\mu\nu}$ is  the quadratic term in the energy-momentum tensor arising
from the extrinsic curvature terms in the projected Einstein tensor
\begin{equation}
{\cal S}_{\mu\nu}=\frac{1}{12}{^{(4)}T}{^{(4)}T}_{\mu\nu}-\frac{1}{4}{^{(4)}T}_{\mu\alpha}{^{(4)}T}^\alpha_\nu
+\frac{1}{24}{^{(4)}g}_{\mu\nu}\Bigl[3{^{(4)}T}_{\alpha\beta}{^{(4)}T}^{\alpha\beta}-{^{(4)}T}^2\Bigr]. \label{eqn12}
\end{equation}
The second correction term ${\cal E_{\mu\nu}}$ is given by
\begin{equation}
{\cal E}_{\mu\nu}={^{(5)}C}_{\alpha\gamma\beta\delta}n^\gamma n^\delta {^{(4)}g}_\mu^\alpha {^{(4)}g}_\nu^\beta, \label{eqn13}
\end{equation}
and is a part of the 5D Weyl tensor and carries information of the gravitational field outside the brane and is constrained by the motion of the matter on the brane, i.e., the Codazzi equation
\begin{equation}
{^{(4)}\nabla}_\mu K^\mu_\nu-{^{(4)}\nabla}_\nu K={^{(5)}R}_{\mu\rho}{^{(4)}g}_\nu^\mu n^\rho. \label{eqn14}
\end{equation}
Further, we have for the extrinsic curvature from the junction conditions
\begin{equation}
K_{\mu\nu}=-\frac{1}{2}\kappa_5^2\Bigl({^{(4)}T}_{\mu\nu}+\frac{1}{3}(\Lambda_4-{^{(4)}T}) {^{(4)}g}_{\mu\nu}\Bigr), \label{eqn15}
\end{equation}
and the 4D contracted  Bianchi equations
\begin{equation}
{^{(4)}\nabla^\nu} {{\cal E}}_{\mu\nu}=\kappa_5^4{^{(4)}\nabla^\nu} {{\cal S}}_{\mu\nu}. \label{eqn16}
\end{equation}
From the 5D Einstein and Bianchi equations one obtains supplementary equations
\begin{equation}
{\cal L}_n K_{\mu\nu}=K_{\mu\alpha}K^\alpha_\nu-{\cal E}_{\mu\nu}-\frac{1}{6}\lambda_5 {^{(4)}g}_{\mu\nu}, \label{eqn17}
\end{equation}
\begin{eqnarray}
{\cal L}_n {\cal E}_{\mu\nu}={^{(4)}\nabla}^\alpha {\cal B}_{\alpha(\mu\nu)} +\frac{1}{6}\Lambda_5( K_{\mu\nu}-{^{(4)}g}_{\mu\nu}K)+ K^{\alpha\beta} R_{\mu\alpha\nu\beta} \cr
+3K^{\alpha}_{(\mu} {\cal E}_{\nu )\alpha}-K{\cal E}_{\mu\nu} +(K_{\mu\alpha}K_{\nu\beta}-K_{\alpha\beta}K_{\mu\nu})K^{\alpha\beta}, \label{eqn18}
\end{eqnarray}
\begin{equation}
{\cal L}_n {\cal B}_{\mu\nu\alpha}=-2{^{(4)}\nabla}_{[\mu}{\cal E}_{\nu]\alpha}+K_\alpha ^\beta {\cal B}_{\mu\nu\beta}-2{\cal B}_{\alpha\beta[\mu}K_{\nu]}^\beta, \label{eqn19}
\end{equation}
with ${\cal L}_n$ the Lie derivative with respect to $n^\mu$ and ${\cal B}$ the "magnetic" part of the bulk Weyl tensor, ${\cal B}_{\mu\nu\alpha}={^{(4)}g}_\mu^\tau {^{(4)}g}_\nu^\sigma {^{(4)}g}_\alpha^\lambda {^{(5)}C}_{\tau\sigma\lambda\epsilon}n^\epsilon$.
From these supplementary equations one can also separate the equations in Eq. (\ref{eqn7}) and in Eq. (\ref{eqn9}) for $W_1(t,r)$ and $W_2(y)$ which proves that the supplementary equations are consistent with the bulk equations.
From the 4D scalar-gauge field equations we obtain
\begin{equation}
\ddot{P}-P''=-\frac{P'}{r}+2(P'\psi'-\dot{P}\dot\psi)- \epsilon^2 W_1^2 e^{2\gamma -2\psi}PX^2, \label{eqn20}
\end{equation}
\begin{equation}
\ddot{X}-X''=\frac{X'}{r}+2\Bigl(\frac{W_1' X'}{W_1}-\frac{\dot{W_1} \dot{X}}{W_1}\Bigr)-\frac{e^{2\gamma}XP^2}{r^2}-\frac{1}{2}W_1^2 e^{2\gamma-2\psi}\beta X(X^2-\eta^2), \label{eqn21}
\end{equation}
which are consistent with those  obtained by Gregory \cite{Greg:1989}.
From the Einstein equations we obtain
\begin{eqnarray}
\dot{\gamma}=\frac{1}{\dot{W_1}}\Bigl[\frac{W_1'}{r}-\gamma ' W_1'-\frac{1}{2r}\gamma 'W_1 -\frac{1}{2W_1}(W_1'^2+3\dot{W_1}^2)+\frac{1}{2}W_1(\dot{\psi}^2+\psi '^2) \cr
+W_1''+W_1'\psi '+\dot{W_1}\dot{\psi} +\frac{3}{8}\kappa_4^2\Bigl(\frac{e^{2\psi}(\dot{P}^2+P'^2)}{\epsilon ^2r^2W_1}+W_1(X'^2+\dot{X}^2)\Bigr)\cr
+\frac{\kappa_5^4}{128}\Bigl(\beta(X^2-\eta^2)^2
+8\frac{e^{2\psi}X^2P^2}{r^2W_1^2}+4\frac{e^{2\psi-2\gamma}(\dot{X}^2-X'^2)}{W_1^2}\Bigr)\cdot\cr
\Bigl(\frac{e^{2\psi}(\dot{P}^2+P'^2)}{\epsilon ^2r^2W_1}+W_1(X'^2+\dot{X}^2)\Bigr)\Bigr], \label{eqn22}
\end{eqnarray}
\begin{eqnarray}
\ddot{\psi}-\psi ''=\frac{1}{r}\psi '-\frac{W_1'}{rW_1}+\frac{2}{W_1}(W_1'\psi'-\dot{W_1}\dot{\psi})+\frac{3\kappa_4^2}{4r^2}\Bigl(
\frac{e^{2\psi}(\dot{P}^2-P'^2)}{\epsilon^2W_1^2}-X^2P^2e^{2\gamma}\Bigr)\cr+\frac{\kappa_5^4}{64r^2}\Bigl(\beta(X^2-\eta^2)^2-4\frac{e^{2\psi}X^2P^2}{r^2W_1^2}
-8\frac{e^{2\psi-2\gamma}(\dot{X}^2-X'^2)}{W_1^2}\Bigr)\cdot\cr
\Bigl(\frac{e^{2\psi}(\dot{P}^2-P'^2)}{\epsilon ^2W_1^2}-X^2P^2e^{2\gamma}\Bigr). \label{eqn23}
\end{eqnarray}
In the 4D case\cite{Greg:1989}, assuming $r_{CS}<< R_H$, one can factor out a  time-dependent "scale"-factor in the 4D FLRW spacetime $U(t)$ in the metric $ds^2=U(t)^2[e^{2(\gamma -\psi)}(-dt^2+dr^2)+e^{2\psi}dz^2+r^2e^{-2\psi}d\varphi^2]$ and
terms in the field equations of the form $\dot{U}/U$, could safely omitted. The overall average cosmological expansion is not affected by the string.
In our case, this is not allowed. One proves\cite{Greg:1989} that in the 4D case the ratio $\dot U/U\approx r_{CS}/R_H$. In our warped 5D case one cannot compare $\dot W_1/W_1$ with $\dot{U}/U$.
It is determined by Eq.(\ref{eqn10}) and will depend on the parameters of the cosmic string, for example the mass per unit length $G\mu\approx \eta^2$.
By this  warpfactor $W_1$, one obtains a huge angle deficit, or equivalently, a tremendous mass per unit length $G\mu\sim 1$ , while the warped manifestation in the brane will be warped down to GUT scale, consistent with observations.

The time dependent part of the warpfactor  causes disturbances of the order much larger than the expected values in the 4D case.
We observe that the several terms on the righthand side of the metric field equation  Eq. (\ref{eqn23}) are  not influenced by the $\pm$ sign of the two branches of the warpfactor. On the righthand side of  Eq. (\ref{eqn22}) we have in the denominator the overall term $\dot{W_1}$. So for an extremum of $W_1$, the solution diverges. This could have a significant influence on a transition from acceleration to a deceleration or vice versa. A reflecting point [ see figure 1] does not cause this problem.

\section{Numerical Solution}
We solved numerically the set of PDE's and plotted in figure 2 a typical solution for the choice of $W_1= 0.1\sqrt{\frac{e^{r+t}}{r}}$.
\begin{figure}[pb]
\centerline{
\includegraphics[width=4.5cm]{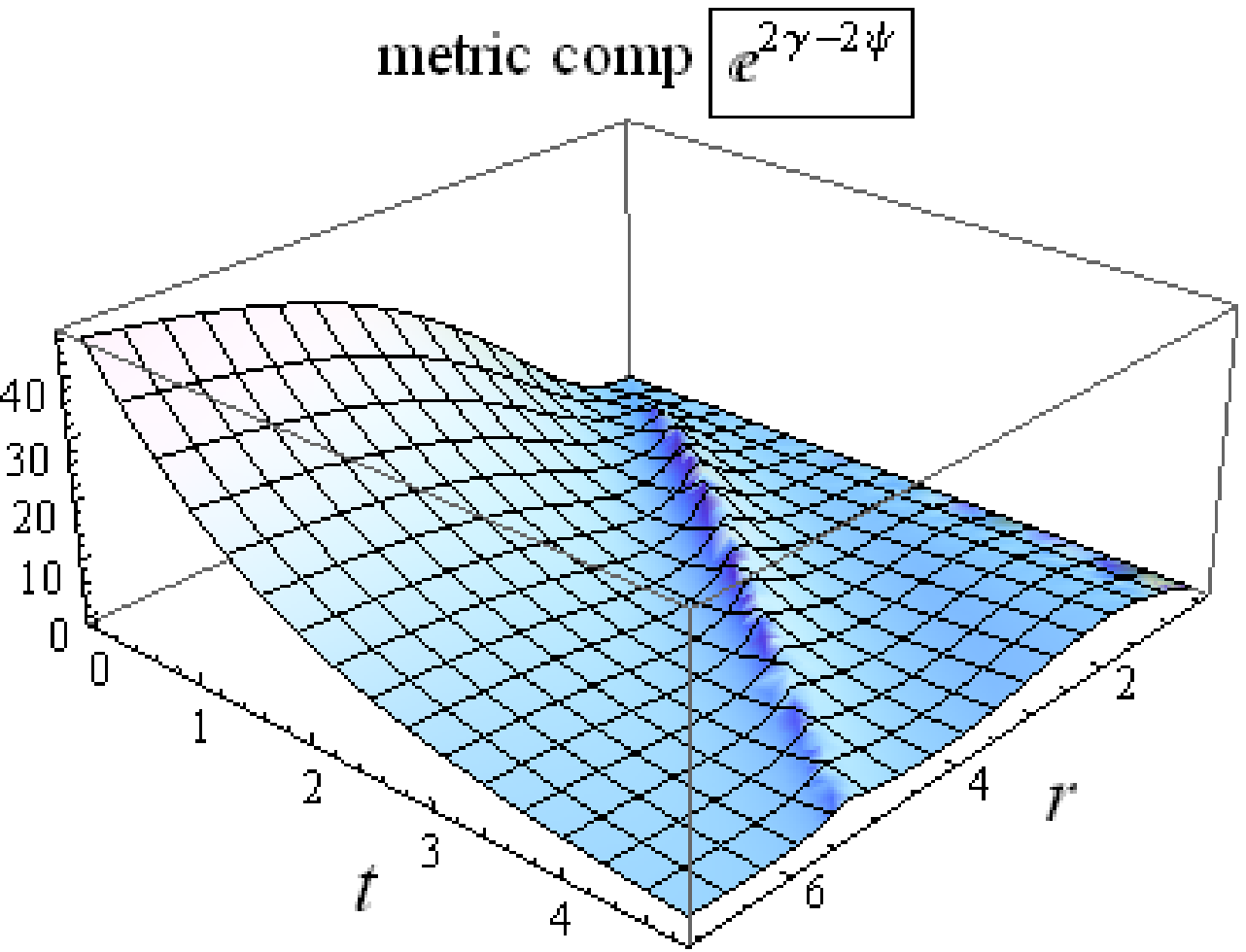}
\includegraphics[width=4.5cm]{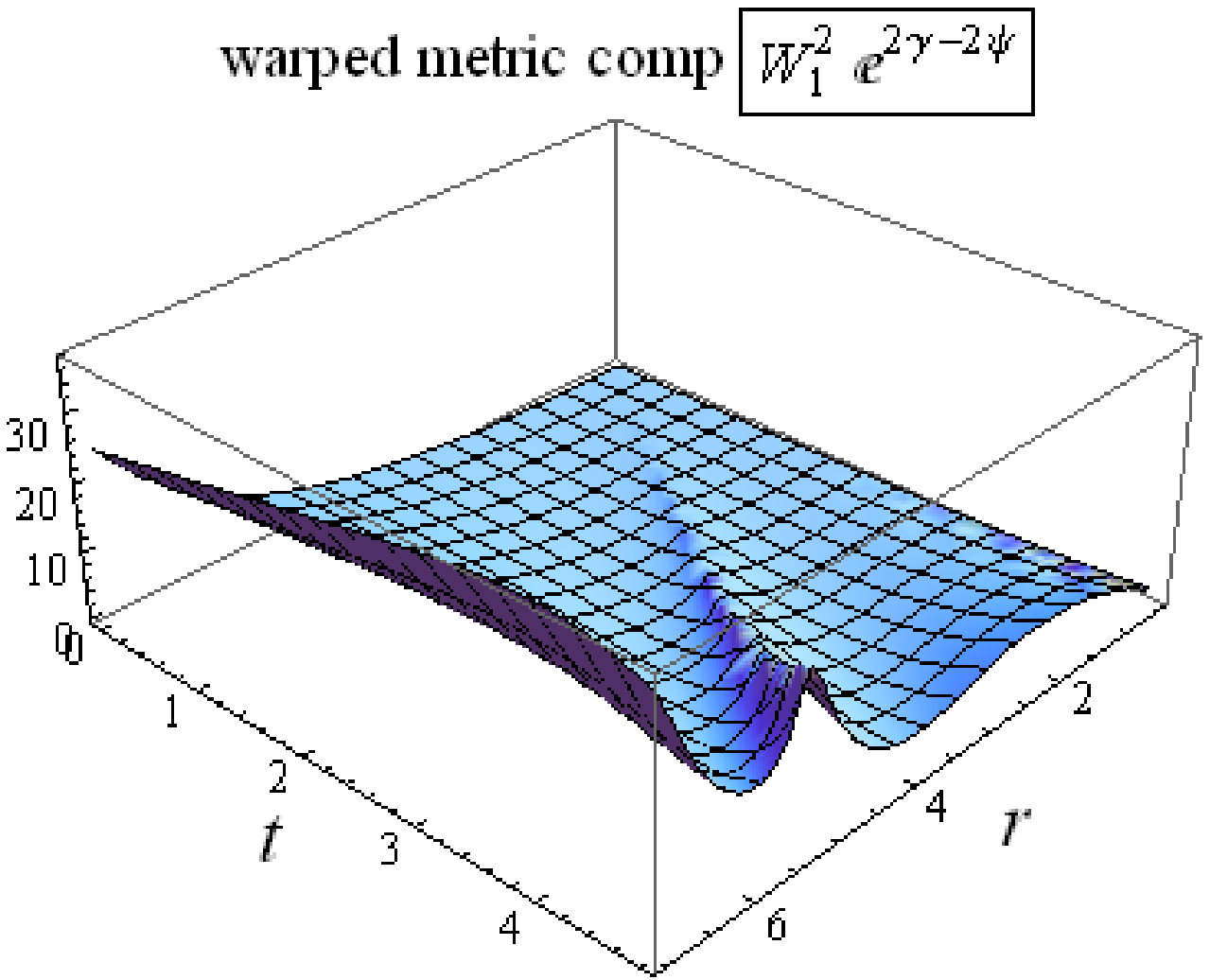}
\includegraphics[width=4.5cm]{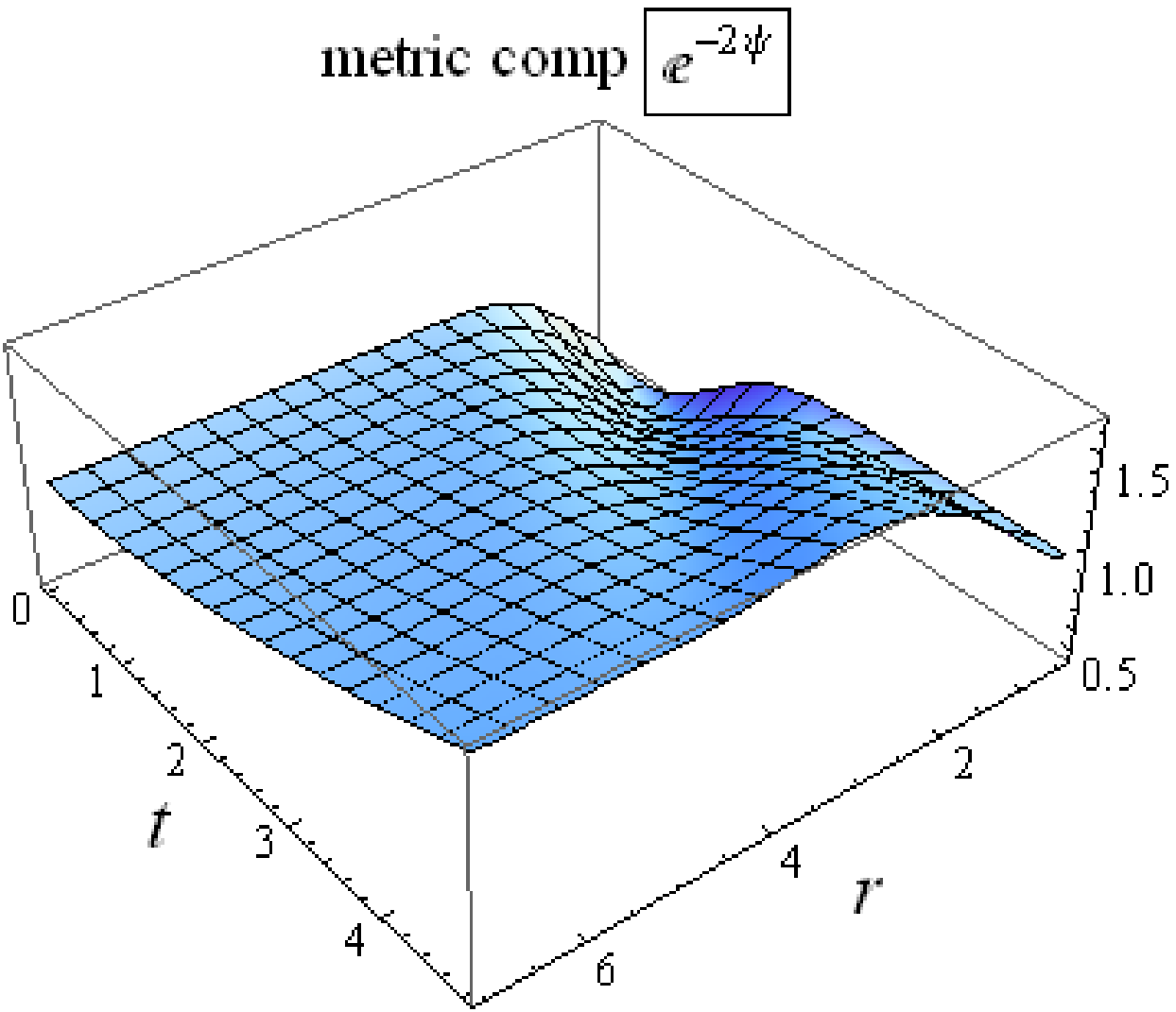}}
\vskip 0.4cm
\centerline{
\includegraphics[width=4.5cm]{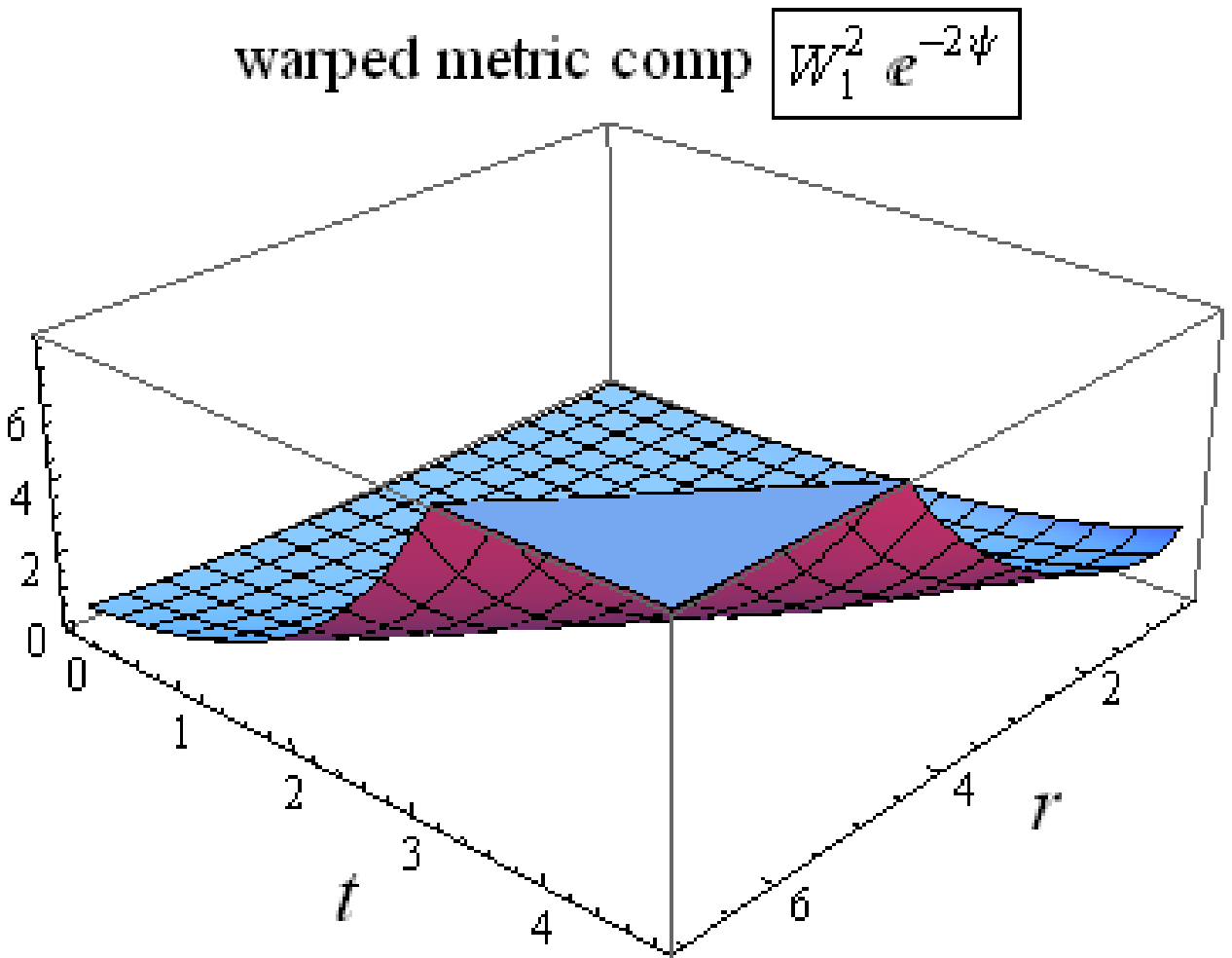}
\includegraphics[width=4.5cm]{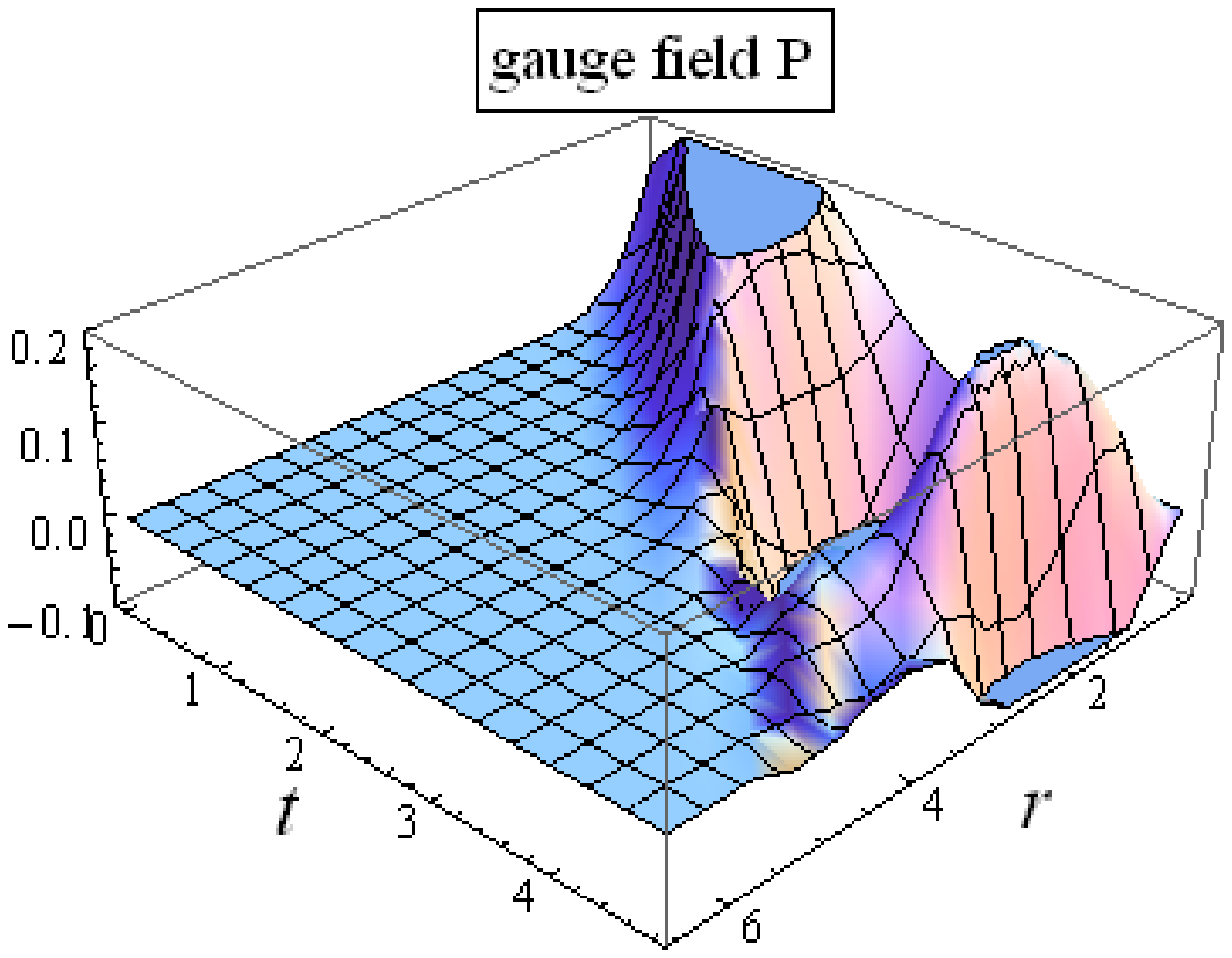}
\includegraphics[width=4.5cm]{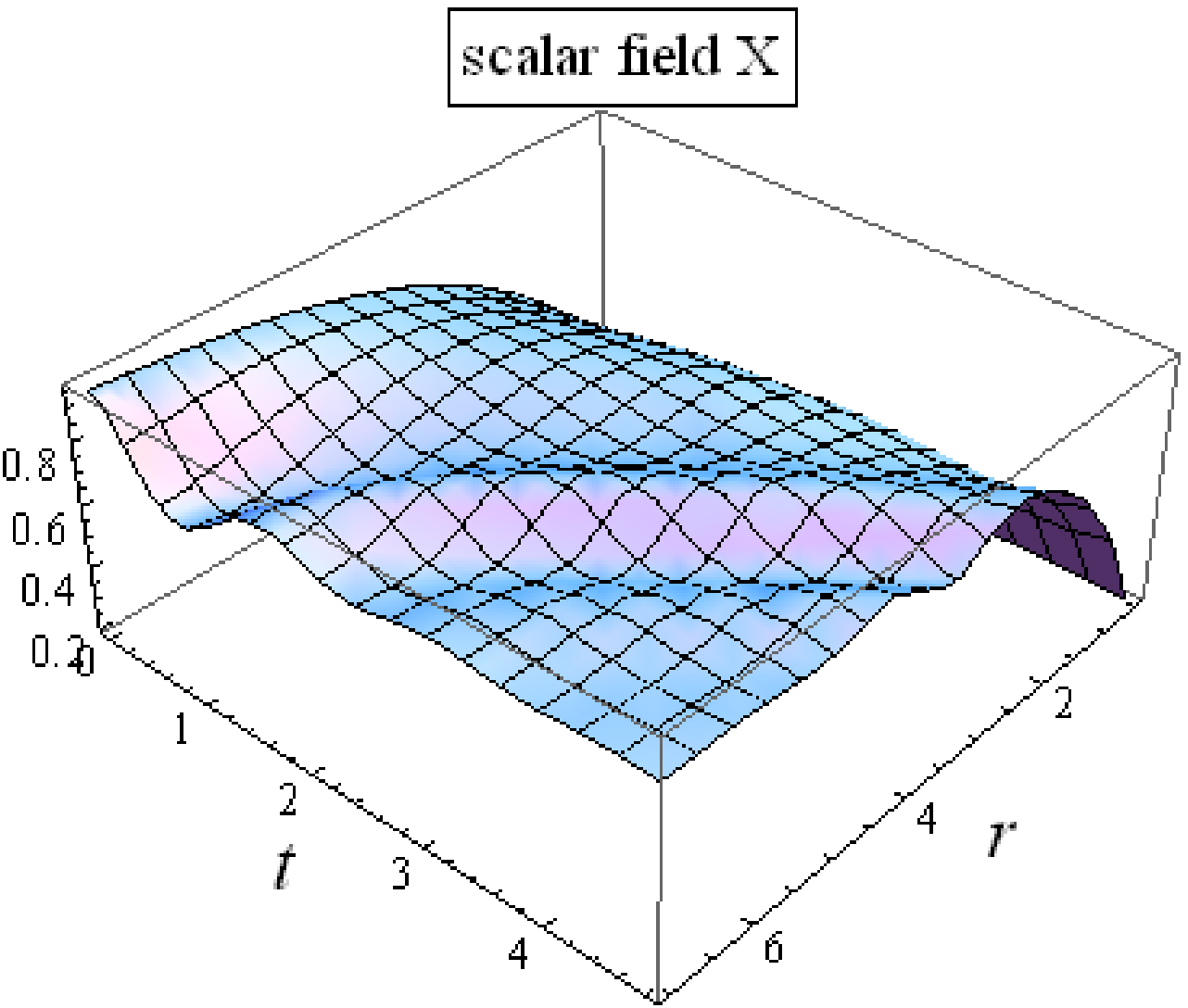}}
\vskip 0.4cm
\centerline{
\includegraphics[width=4.5cm]{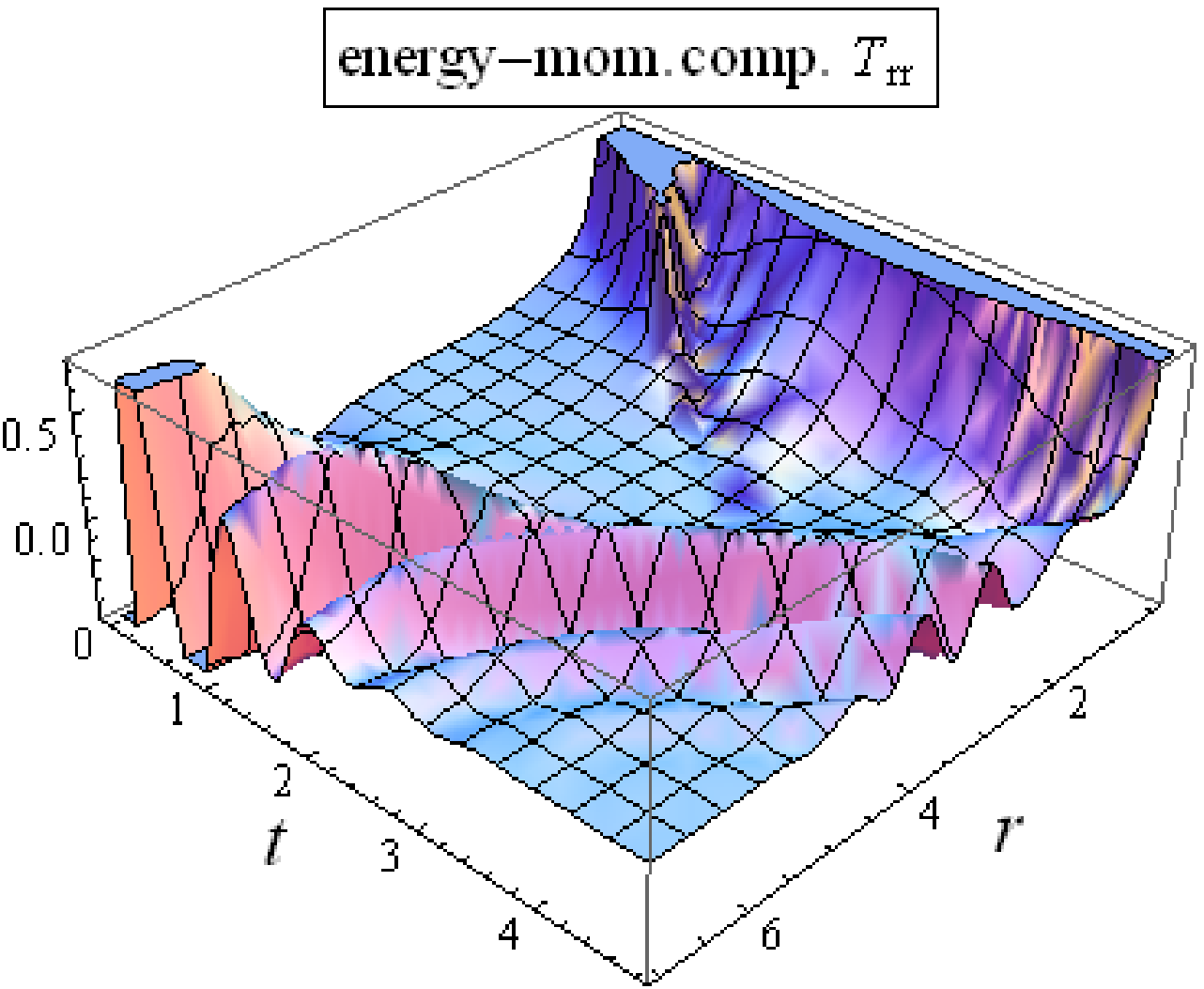}
\includegraphics[width=4.5cm]{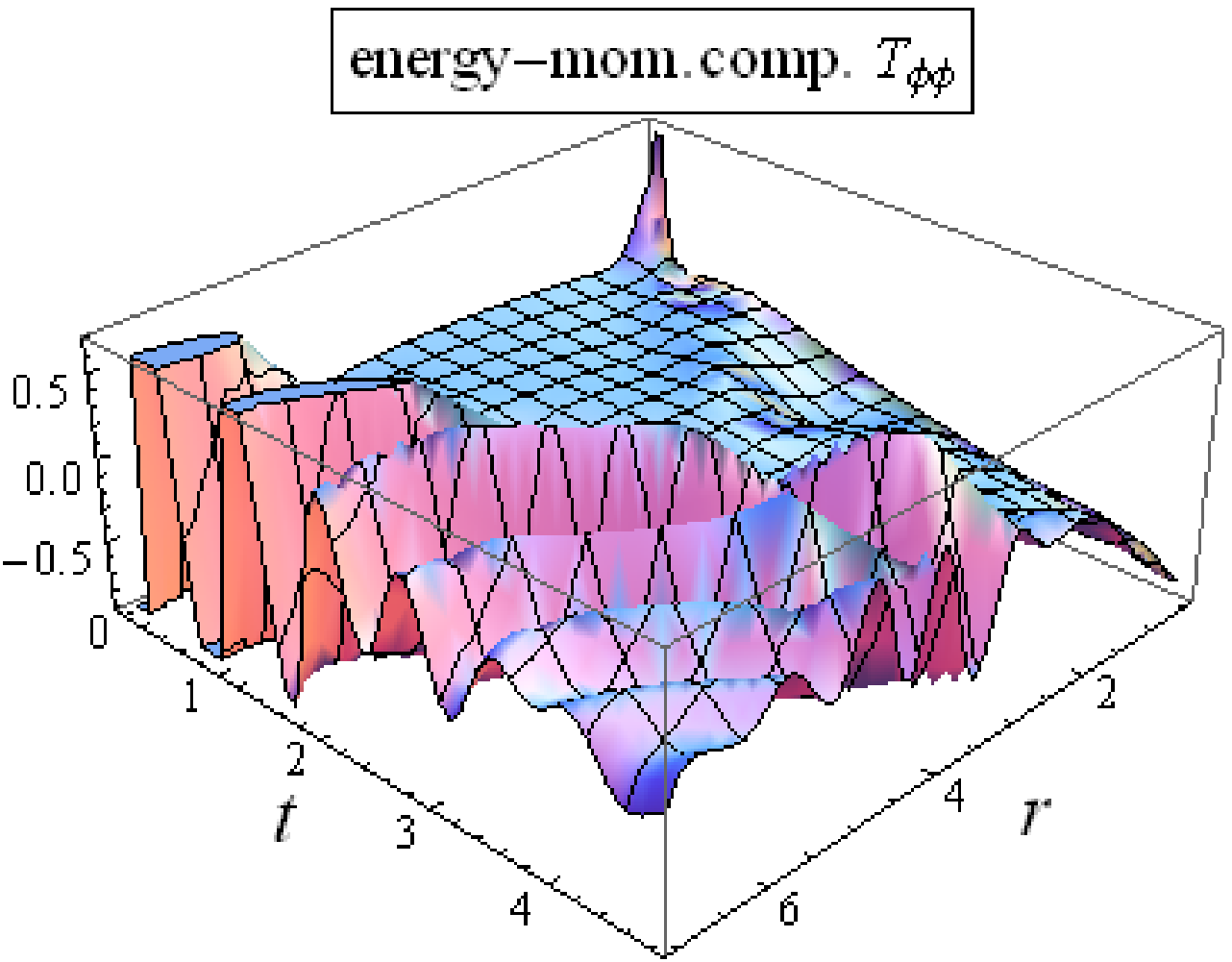}
\includegraphics[width=4.5cm]{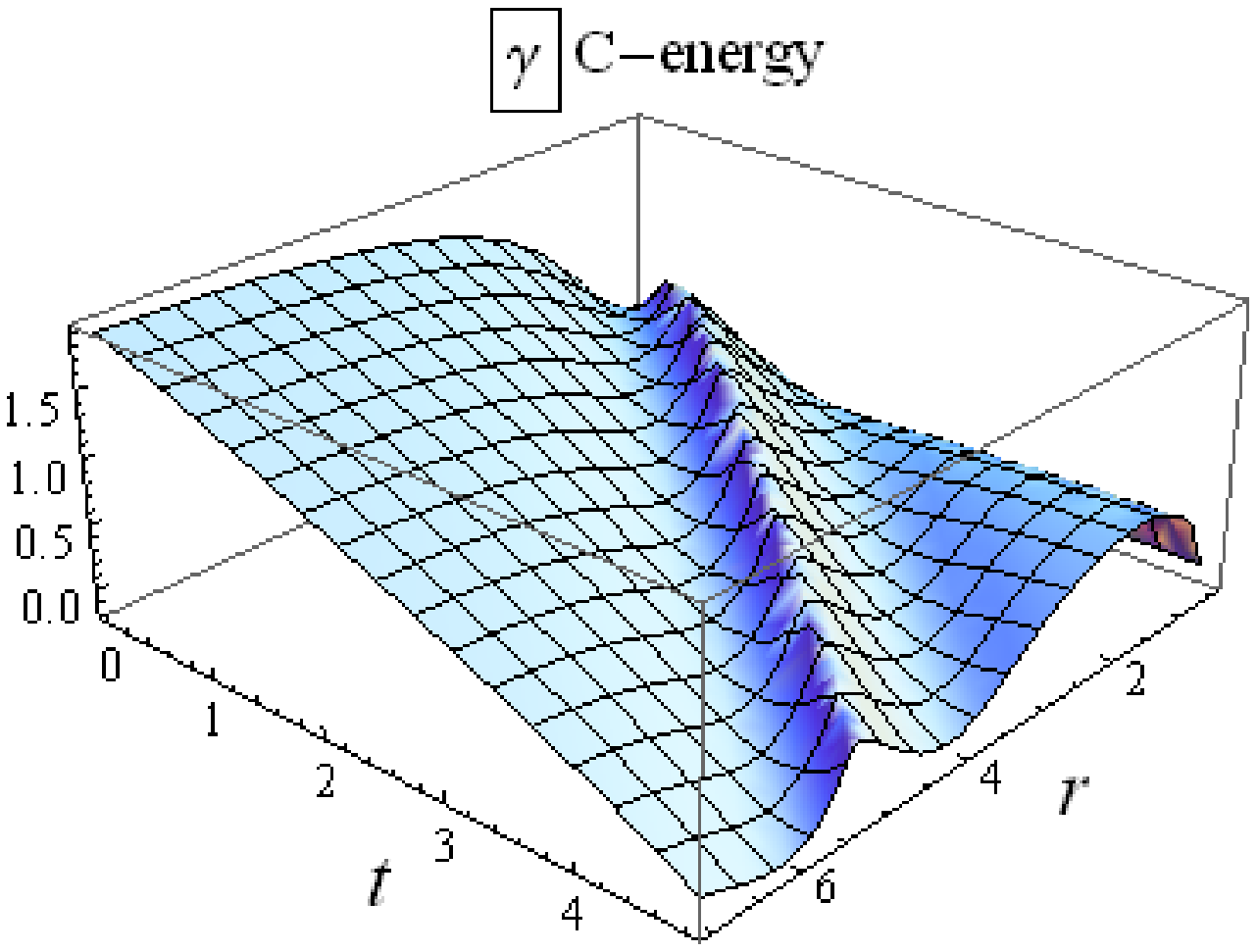}}
\vskip 0.4cm
\centerline{
\includegraphics[width=4.5cm]{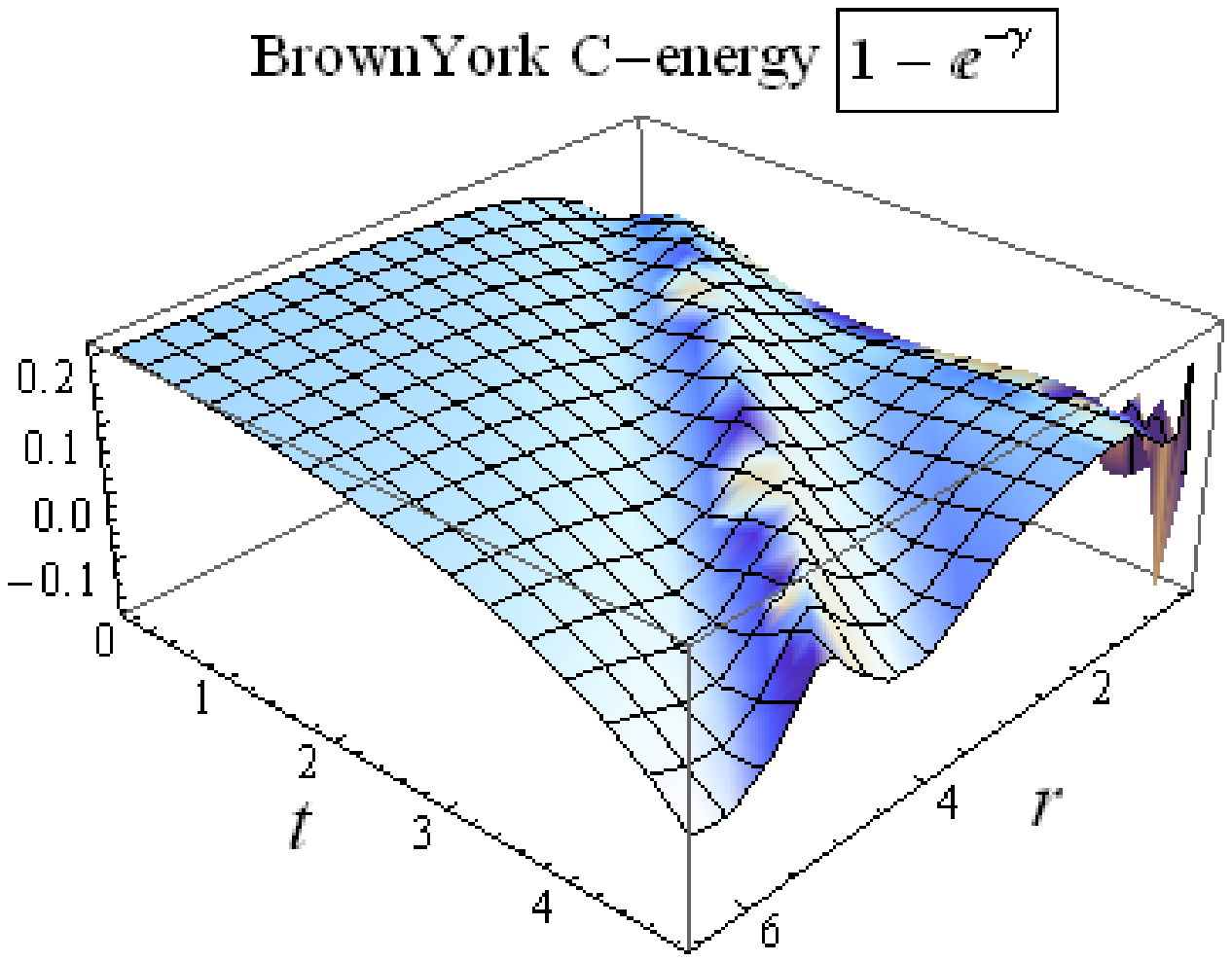}
\includegraphics[width=4.5cm]{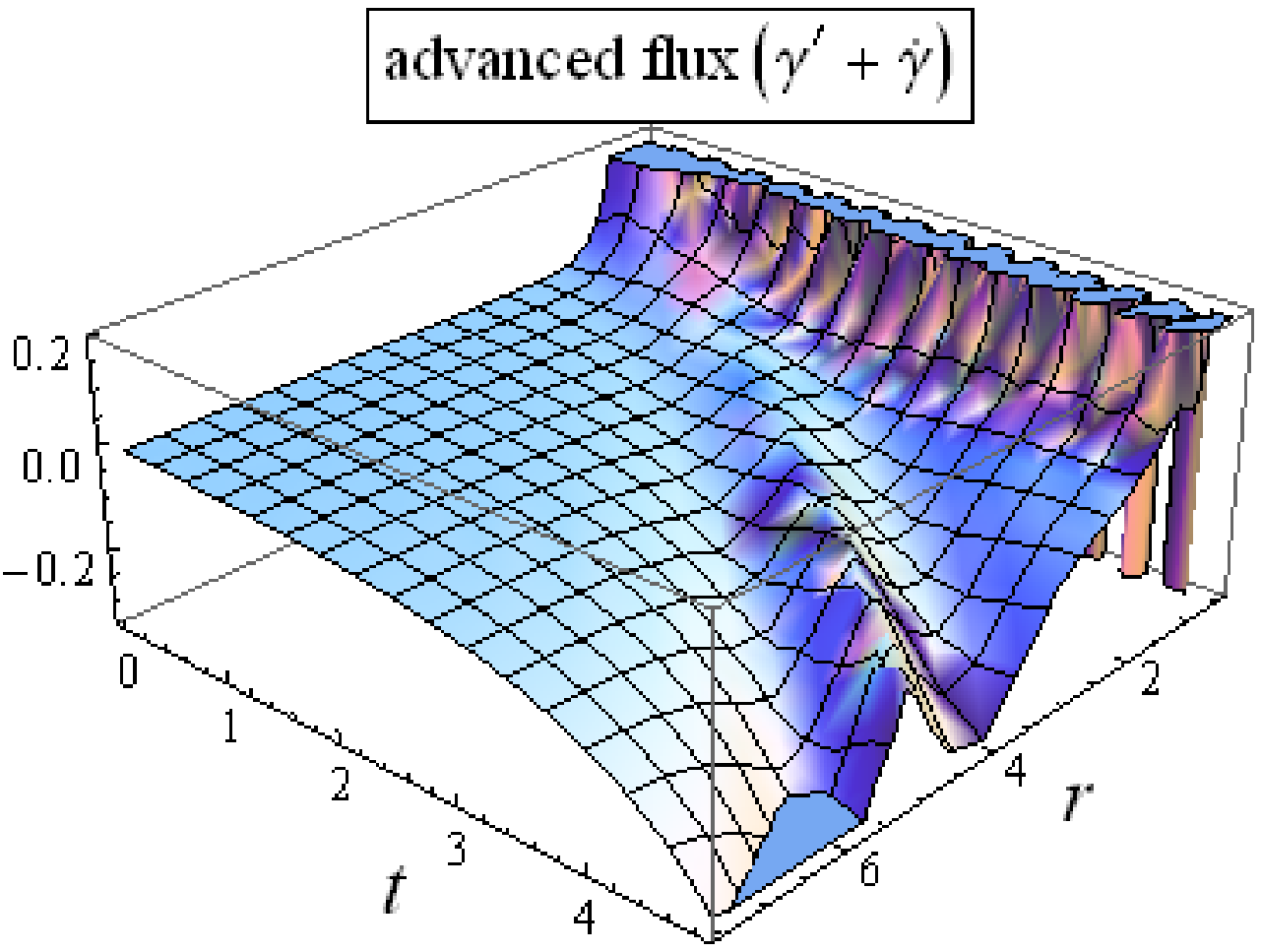}
\includegraphics[width=4.5cm]{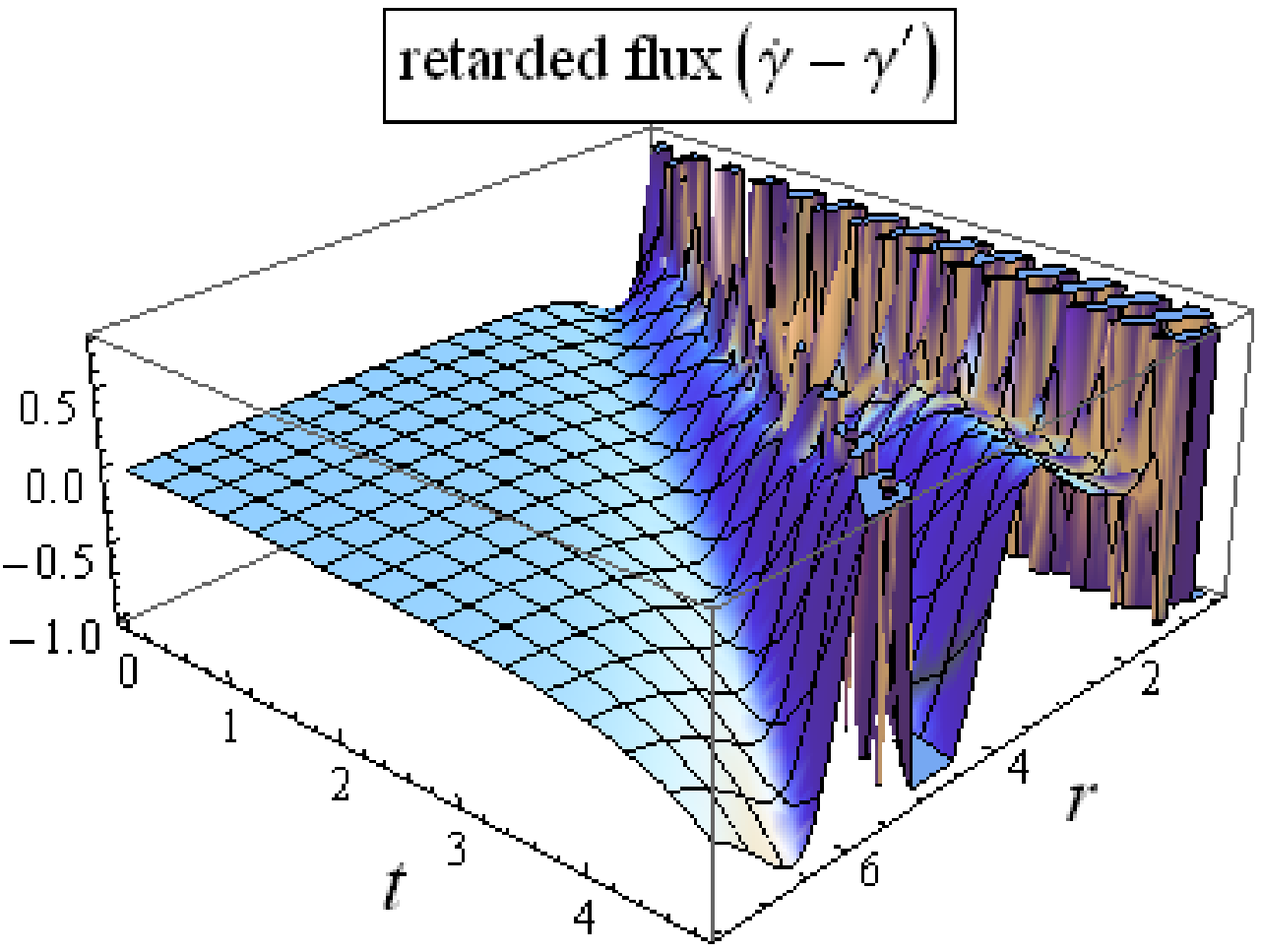}}
\vskip 0.4cm
\centerline{
\includegraphics[width=4.5cm]{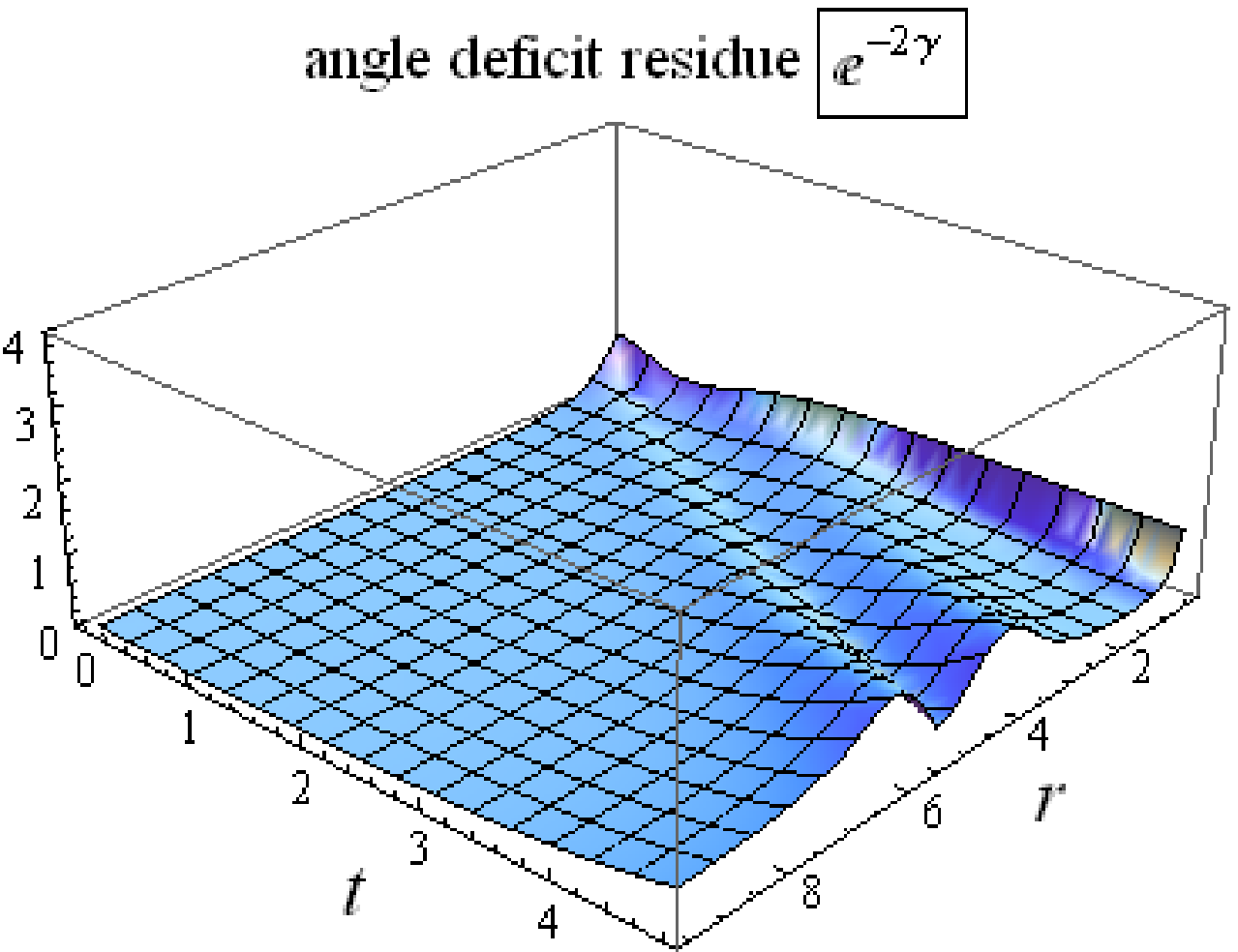}
\includegraphics[width=4.5cm]{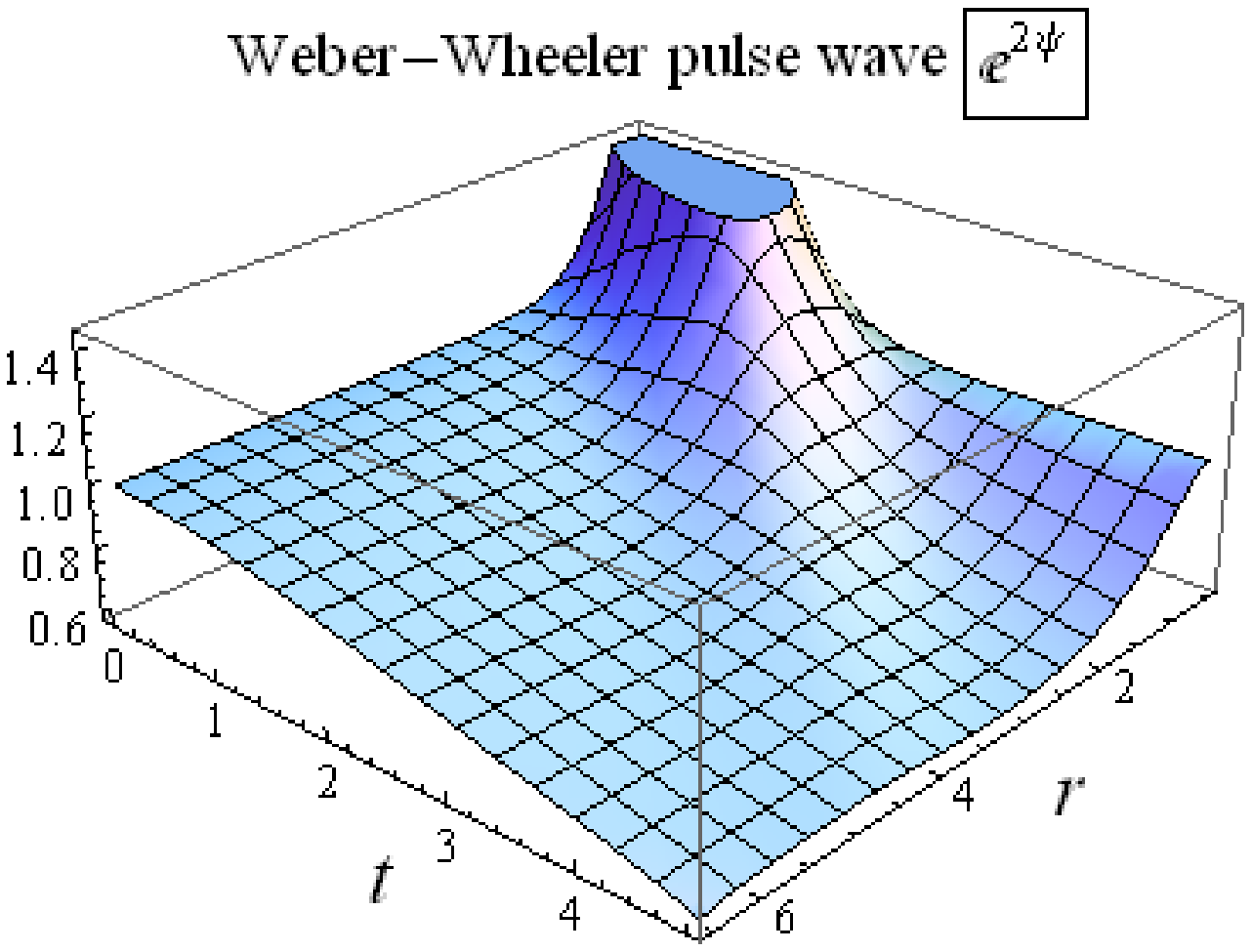}}
\vskip 0.4cm
\caption{Typical solution of the metric components $e^{2\gamma-2\psi}$ and $e^{-2\psi}$ and their warped counterparts. Further, the scalar and gauge fields, the effective energy-momentum tensor components ${^{(4)}T}_{rr}$ and ${^{(4)}T}_{\varphi\varphi}$ and the  C-energy $\gamma$. We also plotted the Brown-York quasi-local C-energy  $1-e^{-\gamma}$ and the advanced and retarded  fluxes $\dot\gamma +\gamma'$ and $\dot\gamma -\gamma'$ respectively. The last two plots are the angle deficit residue $e^{-2\gamma}$ and the Weber-Wheeler pulse wave. }
\end{figure}
We used value-boundary conditions at $r= r_0$ and Neumann boundary conditions  at $r= r_{end}$. For the initial values we choose: $\psi(0,r)= \frac{1}{2}e^{-2r}\sin{3r}$ (Weber-Wheeler pulse wave), $\gamma(0,r)=0$, $P(0, r)= e^{-0.5r}$ and $X(0, r)= 1-e^{-0.5r}$.
It is evident that the emission of retarded wave energy of the scalar-gauge field, triggered by $W_1$, changes the evolution of the induced brane metric. This is a non-local effect from the bulk. The relevant components of the energy-momentum tensor ${^{(4)}T}_{\mu\nu}$ are $T_{rr}$ and $T_{\varphi\varphi}$:
\begin{equation}
{^{(4)}T}_{rr}=-\frac{1}{8}\beta W_1^2e^{2\gamma -2\psi}(X^2-\eta^2)^2+\frac{1}{2}(\dot X^2+X'^2 )+\frac{1}{2}\frac{e^{2\psi}}{\epsilon^2 r^2W_1^2}(\dot P^2+P'^2 )-\frac{1}{2}\frac{X^2P^2e^{2\gamma}}{r^2}, \label{eqn24}
\end{equation}
\begin{equation}
{^{(4)}T}_{\varphi\varphi}=-\frac{1}{8}\beta W_1^2e^{-2\psi}r^2(X^2-\eta^2)^2+\frac{1}{2}e^{-2\gamma}r^2(\dot X^2-X'^2) -\frac{1}{2}\frac{e^{2\psi-2\gamma}}{\epsilon^2 W_1^2}(\dot P^2-P'^2 )+\frac{1}{2}X^2P^2. \label{eqn25}
\end{equation}
We observe that there are two terms which will become dominant for different epochs of cosmic time, i.e., the  potential term $(X^2-\eta^2)^2$, acting as a tension  and the gauge field term $(P_t^2 \pm P_r^2)$. They both contain  the $W_1^2$-term in the numerator and denominator respectively. The same arguments hold for the quadratic term ${\cal S}_{\mu\nu}$.
The pure gauge field term $(\dot P^2\pm P'^2)$ in $T_{rr}$ and $T_{\varphi\varphi}$ will oscillate in time but can change sign in $T_{\varphi\varphi}$, so alternate between pressure (plus sign) and tension (minus sign). In the static case,  $T_{rr}$ and $T_{\varphi\varphi}$ will only contain the term $P'^2$ and is positive, meaning that work is required to squeeze the lines of the magnetic field toward the axis in the radial direction. In the case of a negative sign one would need work in order to stretch the lines of the magnetic field.
From the numerical solution of figure 2 we observe that $T_{\varphi\varphi}$  can become negative. In the 4D case\cite{Garf:1985}, this change of sign of $T_{\varphi\varphi}$ is initiated by the change of $\alpha$( the gauge to scalar mass  ratio) from a value $>1$ to  a value $<1$. Here we find a dynamical change.
The same arguments hold for the pure scalar term $(\dot X^2\pm X'^2)$. However, it doesn't contain the warpfactor.

In figure 2 we also plotted the C-energy $\gamma(t,r)$ (abbreviation for "cylindrical" energy introduced by Thorne\cite{Thorne:1965}), the Brown-York quasi-local C-energy $C_{BY}\equiv(1-e^{-\gamma})$\cite{Gon:2003} and the advanced and retarded C-energy flow along the two null-directions, ($\dot\gamma+\gamma '$) and ($\dot\gamma-\gamma '$).
The C-energy was originally introduced by Thorne as a very useful tool in analyzing the dynamics of a whole-cylinder-symmetric system (i.e., cylindrical symmetry with hyper-surface orthogonal Killing fields) and to describe the nature of Einstein-Rosen gravitational waves and electro-magnetic waves. It was useful to demonstrate the resistance of magnetic-field lines to cylindrically gravitational collapse and to discuss the dynamics of Melvin's magnetic universe when large radial perturbations are introduced\cite{Thorne:1966}.
Evidence for a bona fide "energy" status of $\gamma$ is the fact that its coordinate gradient is related to a flux vector field which is covariantly conserved\cite{Gon:2003}.
It was found that the Brown-York quasi-local C-energy description is a better measure of the C-energy than the original one of Thorne. $C_{BY}$ is derived by means of a Hamiltonian reduction of the field equations, wherein the $C_{BY}$-energy arises naturally from the Hamiltonian constraint, while Thorne's arguments were merely heuristic.
Another argument for the choice of $C_{BY}$ is the fact that the asymptotic metric in the vacuum case can be written in the form
\begin{equation}
ds^2=-dt^2+dr^2+dz^2+(1-\bar C_{BY})^2r^2d\varphi^2,\label{eqn26}
\end{equation}
with $\bar C_{YB}$ the asymptotic value for $r\rightarrow\infty$. This can be seen by the re-scaling $t\rightarrow e^{-\gamma_\infty +\psi_\infty} t$, $r\rightarrow e^{-\gamma_\infty +\psi_\infty}r$ and $z\rightarrow e^{-\psi_\infty}z$ in the asymptotic form of metric of (\ref{eqn5}). This metric is exactly the metric exterior to a cosmic string with mass per unit length $\mu\equiv \bar C_{BY}$. This is only valid in the weak-field limit $\mu <<1$.
Because $\gamma$ doesn't decouple from the equation for $\psi$, we have two degrees of freedom, i.e., $\psi$ and $\gamma$,  for the gravitational waves (in the Einstein-Rosen case there is one degree of freedom, because $\gamma$ can be obtained by quadrature when $\psi$ is known). Further we have the warpfactor $W_1$ as extra metric component.
It turns out that the wavelike behavior depends critically on the ratio of the scalar to gauge masses, $\alpha \equiv \sqrt{\frac{m_P}{m_X}}$. We took $\alpha >>1 $.
For $\alpha <<1$ one obtains the features of a global string, which is in general singular\cite{Slag2:2014}.
From the plots of figure 2 we can draw some global conclusions concerning regular  behavior, the interaction of de cosmic string with the gravitational waves  and the deviation from the 4D FLRW counterpart model ( with the same matter field and cylindrical symmetry).
The C-energy is radiated along the null directions in agreement with other research on cylindrical symmetric spacetimes. Although our numerical code runs out of accuracy for larger r and t values, we observe a fall-off of the Weber-Wheeler pulse $e^{2\psi}$.
The induced disturbances in the scalar field X travel backwards closer to the axes of the cosmic string and the scalar field returns to its original static form with its core closer to r=0.

Some remarks must be made with regards to the asymptotic behavior of the metric Eq.(\ref{eqn26}). If we define $\varphi '=(1-\bar C_{YB})\varphi$, then the metric  Eq.(\ref{eqn26}) becomes Minkowski. However, $\varphi '$  now takes the values $0\leq\varphi ' \leq (1-\bar C_{YB})2\pi$. So if $\bar C_{YB}<1$, we obtain a angle deficit and the spacetime is conical. Because $\bar C_{YB}$ is related to the metric component $e^{-2\gamma_{\infty}}$\cite{Gon:2003}, we can plot this angle deficit residue $e^{-2\gamma}$
and compare its behavior with the 4D counterpart model of Gregory\cite{Greg:1989}. In the static 4D case, so in the absence of radiation, the angle deficit for isolated strings at large radial distances is related to the mass per unit length $\mu$ of the cosmic string, i.e., $\Delta\theta = 8\pi G\mu$. In the weak field approximation one has $\mu\sim\eta^2$. So for GUT strings ($\eta <10^{-2}$) we expect that the asymptotic behavior is practically Minkowski minus a wedge. The angle deficit will also depend on $\alpha$. A typical value for the angle deficit is $\Delta\theta \approx 2. 10^{-3}$ for $\alpha =64$\cite{Lag1:1987}.
If one admits radiative effects on the 4D time-dependent spacetime, one proves\cite{Greg:1989} that the string-cosmology spacetime  essentially looks like a scaled version of a cosmic string solution,  not desirable (see Eq. (\ref{eqn26})) because we live now in a flat FLRW spacetime. In deriving this result, one makes the simplification that the gravitational wave energy decouples from the cosmic string energy.
The energy of the string is then exponentially damped and one finds that asymptotically $e^{-2\gamma}$ remains constant for all time.
In our 5D warped model, we made no simplifications, because we solved the full set of coupled wave equations numerically. Moreover, we have the contribution from the bulk as explained in section (2.2).
We observe from figure 2 that $e^{-2\gamma}$ increases with time.
One can prove\cite{Lag2:1989}, that for increasing $e^{-2\gamma}$, the angle deficit becomes $>2\pi$ and one deals with a Kasner-like spacetime.
So we conclude that in our warped cosmological  model we don't deal with a scaled version of the standard cosmic string. The notion of a conical spacetime fades away on our effective ${^{(4)}g}_{\mu\nu}$ spacetime.
For values of $\eta >>10^{-2}$ above the GUT scale, gravitational effects become more relevant and the weak-field approximation can no longer be applied. These so-called supermassive cosmic strings possess in 4D, however, a curvature singularity\cite{Lag2:1989}. Moreover, they probably would have  formed before an inflationary era and subsequently inflated away. In our warped 5D spacetime  these supermassive cosmic strings arise in a different manner, i.e., by the warpfactor.
We can also compare  the plots of $W_1^2 e^{-2\psi}$ and $e^{-2\psi}$ of figure 2, in order to draw some conclusions about the asymptotic conical behavior.
The metric component $g_{\varphi\varphi}=e^{-2\psi} r^2$ approaches in the classical 4D cosmic string solution to the value $(1-4G\mu)^2 r^2$, so in the weak field approximation one expects asymptotically that
$e^{-2\psi_\infty}$ approaches a constant value and $g_{\varphi\varphi}=e^{-2\psi} r^2$ will show a  small deviation from $r^2$. This constant value is dependent of the parameters $\alpha , \beta $ and $\eta$
of the model.
However not in our case. We observe that the warped counterpart metric component $W_1^2e^{-2\psi}$ behavior is different: it increases for larger values of r and t and seems to be  insensitive for the values $\alpha , \beta $ and $\eta$.  So we observe no conical behavior.
This  absence of the angle deficit in a 5 dimensional warped cosmic string model was also found earlier\cite{Slag1:2012}.
If one investigate the brane evolution  via the modified second Friedmann equations, then on the right-hand side  new terms will appear which contain $W_1$, the quadratic contribution of the energy-momentum tensor ( see Eq.(\ref{eqn22}) and (\ref{eqn23})) and a bulk radiation term containing the constants $\tau$ and $c_i, d_i$ ( see  Eq.(\ref{eqn7}) and Eq.(\ref{eqn10})).
From the behavior of $W_1$ of figure 1 and the occurrence of $W_1$ in the denominator of the  evolution equation Eq.(\ref{eqn22})of $\gamma$ one could conclude that an ever lasting exponential expansion is preferable over the other possibilities.
The dependency of our model on the several (integration-) constants is currently under investigation by the authors.
\section{Conclusion}
The late-time epoch of accelerated expansion can be explained within the framework of the 4D standard model of cosmology by hypothesizing a dark energy component to the total energy density of the universe with negative pressure equation of state $p=w\rho$, where the equation of state parameter w lies near -1. Dark energy can possibly be understood as arising  from vacuum energy, but this explanation introduces the cosmological constant problem.
Alternatively, it is possible that there is no dark energy, but instead a low-energy/large scale ( i.e., "infrared") modification to general relativity that accounts for late-time acceleration, schematically, $G_{\mu\nu}+G_{\mu\nu}^{dark}=8\pi G T_{\mu\nu}$.
The higher dimensional graviton has massive 4D modes felt on the brane (KK modes).  Gravity leakage from the 4D brane can cause this self-acceleration without the need to introduce dark energy.

We find a significant new late-time behavior of the warped 5D FLRW cosmological model, when a self-gravitating U(1) scalar field coupled to a gauge field (cosmic string) is present on the brane and the effective cosmological constant on the brane is zero (RS fine-tuning). Compared with the 4D counterpart model, the cylindrically symmetric disturbances  have a significant impact on the late-time expansion rate of the universe. This is caused by the fact that one has no bound on the mass per unit length of the cosmic string, in contrast to the 4D model of Gregory \cite{Greg:1989}. We find an exact solution for the warpfactor, ${\cal W}=W_1(t,r)W_2(y)$.  The time-dependent part, which plays the role of a "scale" factor, is a monotone exponential increasing function with an inflection point or has a extremum.
In the latter case, the metric component $\gamma$ becomes singular at the moment the warp factor develops a extremum. This behavior could have influence on the possibility of a transition from acceleration to  deceleration or vice versa. In the first case the accelerated expansion can be accomplished without the need of an effective brane cosmological constant as dark energy.
The disturbances of the scalar and gauge fields interact with gravitational waves and could act as an effective dark energy field and trigger brane fluctuations.
The scalar field sheds off wave energy and the oscillations slowly decay and eventually the variables return to their static values.

The recently found large-scale alignment of quasar polarizations in our Universe\cite{Hut:2014} could be explained by our model. The modified energy momentum tensor component $T_{zz}$ induces traveling waves in the z-direction on the brane.

\end{document}